\documentclass[aps,prd,preprint,superscriptaddress,nobibnotes,nofootinbib,a4paper]{revtex4-1}
\usepackage[english]{babel}
\usepackage{graphicx}
\usepackage{amsmath,amssymb}
\usepackage[caption=false]{subfig}
\usepackage{setspace}
\usepackage{hyperref}
\usepackage{rotating}

\newcommand{\p}{\partial}

\newcommand{\Tr}{\mathop{\rm Tr}\nolimits}

\newcommand{\beq}{\begin{eqnarray}}
\newcommand{\eeq}{\end{eqnarray}}
\newcommand{\non}{\nonumber\\}

\newcommand{\MeV}{\mathop\textrm{MeV}}

\begin{document}

\title{Loosening up the Skyrme model}

\author{Sven Bjarke Gudnason}
\email{bjarke(at)impcas.ac.cn}
\affiliation{Institute of Modern Physics, Chinese Academy of Sciences,
  Lanzhou 730000, China}

\date{\today}
\begin{abstract}

We consider the Skyrme model with the addition of extra scalar
potentials that decrease the classical binding energies of the
Skyrmions to about the 3\% level -- without altering the pion mass
-- if we insist on keeping platonic symmetries that are usually
possessed by Skyrmions.  
A side effect of the potentials under consideration is the smaller
size of the 1-Skyrmion resulting in a smaller moment of inertia and in 
turn a larger spin contribution to the energy upon semi-classical
quantization.
After taking into account the quantum contributions we find total
binding energies at the 6\% level.

\end{abstract}


\maketitle

\section{Introduction}

The Skyrme model was introduced by Skyrme as a model for baryons in a
low-energy effective field theory of pions
\cite{Skyrme:1961vq,Skyrme:1962vh}. 
It first caught serious attention when it was shown that its soliton
-- the Skyrmion -- is the baryon in the large-$N_c$ limit of QCD
\cite{Witten:1983tw,Witten:1983tx}.
It took, however, a while before the higher-charged Skyrmion solutions
-- corresponding to baryons with $B>2$ -- were found.
The breakthrough came with the introduction of the rational maps,
where the Skyrmion is split into a radial component and a sphere which
is mapped to a Riemann sphere that is wrapped $B$ times
\cite{Battye:1997qq,Houghton:1997kg}. 
The rational maps are believed to describe the minimizers of the
energy functional of the Skyrmions to quite high precision for
vanishing pion mass and for small $B\leq 7$.
For small baryon numbers the pion mass does not have a big impact on
the Skyrmion solutions. However, when the pion mass is turned on -- at
approximately the value of the physical pion mass -- the Skyrmions
prefer to order themselves as a crystal of alpha particles
\cite{Battye:2006na} as opposed to the fullerenes described by the
rational maps.
The Skyrmions do capture many phenomenological features of nuclear
physics and moreover they give a geometrical interpretation of the
physics behind.
Nevertheless, a longstanding problem of the Skyrme model -- which has
been evident from the different calibration attempts
\cite{Adkins:1983hy,Manton:2006tq} -- is that the binding energies
naturally come out too large; about one order of magnitude too large.
More precisely, the recalibration of the Skyrme model in
Ref.~\cite{Manton:2006tq} ameliorates the problem of the large binding
energies by using a higher-charged Skyrmion ($B=6$) as input (as
opposed to the calibration using the proton and delta resonance
\cite{Adkins:1983hy}). 

The problem of too large binding energies has been the motivation for
improving the Skyrme model and gave rise to three recent directions to
do so.
One attempt is to make a model with an infinite tower of mesons,
which is truly BPS in the limit where all the mesons are
included \cite{Sutcliffe:2010et,Sutcliffe:2011ig}. This model is
derived from the self-dual Yang-Mills 
theory in five dimensions. 
The second line of research is based on a modified Lagrangian that is
composed of only a sixth-order derivative term (as opposed to the
standard kinetic term and the fourth-order Skyrme term) as well a
potential; this theory is called the BPS Skyrme model
\cite{Adam:2010fg,Adam:2010ds} and as opposed to the normal Skyrme
model (that does not have solutions saturating its bound), its BPS
bound on the energy can be saturated for solutions with arbitrary
large baryon numbers. 
The third and last attempt to ameliorate the large binding energy was
made from the observation that the pure Skyrme term (fourth-order) as
well as a unique potential to the fourth power saturates an energy
bound \cite{Harland:2013rxa} and thus is BPS for a single baryon
($B=1$). This model is called the lightly bound Skyrme model
\cite{Gillard:2015eia}. Its higher-charged Skyrmions do however not
saturate said bound \cite{Harland:2013rxa}, but they do in fact lie so
close to the bound that the model indeed gives rise to very small
classical binding energies -- of the order of experimental data. 

Although the lightly bound Skyrme model is a promising attempt at
producing viable binding energies for possibly all nuclei, it has a
drastic difference with the normal Skyrme model; namely the shapes of
the Skyrmions \cite{Gillard:2015eia}. Its higher-charged Skyrmion
solutions take the shape of $B$ spheres situated at the vertices of a
face-centered cubic (FCC) lattice. This is quite in contrast to the
Skyrmions of the normal Skyrme model that prefer to sit in a lattice
of alpha particles. The latter is quite a welcomed feature from the
point of view of nuclear clusters \cite{Freer:2007} (see e.g.~Fig.~6
in Ref.~\cite{Freer:2007}), which indeed hint at the importance of the
alpha particles or the $B=4$ solutions in baryons with higher baryon
numbers. 

A remarkable achievement in the Skyrme model is the description of the
Hoyle state in ${}^{12}$C (Carbon-12) and its corresponding band of
rotationally excited states \cite{Lau:2014baa}.
In this normal formulation of the Skyrme model, two classical
solutions with baryon number 12 are found to have almost the same
classical energy, but very different shapes, resulting in moments of
inertia whose ratio is about 2.5 -- in perfect agreement with
experimental data \cite{Lau:2014baa}. The ratio is indeed observable
from the slopes of the rotational bands coming from the ground state
and the Hoyle states, respectively.

The reconciliation of the two above-mentioned results is however hard
to meet. The lightly bound Skyrme model, in contrast to the normal
Skyrme model, predicts twelve spheres situated at the vertices of the 
FCC lattice with nearly the same energies of all its different
configurations (this is of course just a simple argument from the fact
that the overlap of the spheres is marginal and thus the energy is
roughly independent of where the spheres are placed on the nearby
vertices).
It is easy to convince oneself that there are a multiple of different
configurations with almost the same energy, but different moments of
inertia.
This degeneracy is observed already at the classical level for
$B=6,7,8$ in Ref.~\cite{Gillard:2015eia} (for instance, five different
configurations with $B=8$ and nearly the same energy were found) and so
it is expected to be even higher for $B=12$.
Although there might exist one classical Skyrmion configuration with
approximately 7 MeV higher energy than the global minimizer -- the
ground state -- and possibly giving rise to a slope that is 2.5 times
higher than that of the ground state, there will still be too many
other states with different slopes.
Whether quantization or some other mechanism can solve this puzzle is
beyond the scope of the present paper.

The mechanism at work in the lightly bound Skyrme model
\cite{Gillard:2015eia} is a repulsive force due to the nonlinear
potential of the form $(1-\Tr[U]/2)^4$ that acts at short distances
and is strong enough to separate the $B$-Skyrmion into $B$
identifiable spheres that are still bound together.
Notice that due to the nonlinearity of the potential, it does not
alter the linear force present in the Skyrme model without the
addition of this potential. The long-range attractive forces present
in the normal Skyrme model thus remain.
Exactly this type of potential was studied long ago in the baby Skyrme
model \cite{Leese:1989gi}, see also
Refs.~\cite{Hen:2007in,Salmi:2014hsa,Salmi:2015wvi,Samoilenka:2015bsf}.

In this paper the scope is to study (a part of) the parameter space of
a class of potentials
\beq
V_n \propto \frac{1}{n}\left(1 - \frac{1}{2}\Tr[U]\right)^n, \qquad
n>2,
\eeq
exhibiting repulsive forces and determine how
low binding energies can be attained without losing the $B=4$ cube
that is a welcomed feature of the Skyrme model in light of
clustering into alpha particles.
As the parameter space of the linear superposition of several
potentials is obviously huge, we limit ourselves to a slice in the
parameter space spanned by $V_2$ and $V_4$.
$V_4$ is exactly the holomorphic type of potential of the lightly
bound Skyrme model \cite{Gillard:2015eia}, whereas $V_2$ is a similar 
potential with a smaller repulsive force.

We find that both $V_2$ and $V_4$ decrease the classical binding
energies, but $V_2$ is able to lower the classical binding energies
further without breaking the platonic symmetries of the Skyrmions;
however, not quite enough to reach the experimentally observed values
of nuclei.
The inclusion of the pion mass was originally thought to be a minor
effect but its effect is studied over the entire selected region of
parameter space. It turns out that although it lowers the classical
binding energies when the potentials $V_2$ and $V_4$ are turned off,
it actually increases the classical binding energies when a sizable
value of the coefficient of either one of the potentials is turned
on.
Although this effect is less welcome, it also has the effect of
maintaining the platonic symmetries to larger values of the
coefficients of said potentials.
After finding the optimal point in the parameter space -- which turns
out to be at $(m_2,m_4)\sim (0.7,0)$ -- a calibration to physical
units is done and an estimate of the contributions due to spin and
isospin quantization is taken into account.
The result is that the $V_2$ model can retain platonic symmetries and
have total binding energies at the 6\% level (whereas the classical
contribution is near the 3\% level).

The paper is organized as follows. We introduce the Skyrme model with
the additional potentials in Sec.~\ref{sec:model} and present
numerical results in Sec.~\ref{sec:numerical}. Finally, we conclude
with a discussion in Sec.~\ref{sec:discussion}.

\section{The model}\label{sec:model}

The Lagrangian density of the model under study is given by
\beq
\mathcal{L} =
\frac{c_2}{4} \Tr[L_\mu L^\mu] 
+ \frac{c_4}{32} \Tr\left([L_\mu,L_\nu][L^\mu,L^\nu]\right)
- V(U),
\label{eq:L}
\eeq
where $L_\mu\equiv U^\dag\p_\mu U$ is the left-invariant
$\mathfrak{su}(2)$-valued current, $c_2>0$ and $c_4>0$ are
positive-definite real constants, $\mu,\nu=0,1,2,3$ are spacetime 
indices, $U$ is the Skyrme field related to the pions as
\beq
U = \mathbf{1}_2\sigma + i\tau^a\pi^a,
\eeq
obeying $U^\dag U=\mathbf{1}_2$ which translates into
$\sigma^2+\pi^a\pi^a=1$, $\tau^a$ are the Pauli matrices and finally,
the potential is taken to be a function of $\Tr U$ with the vacuum
expectation value of $U$ being at $U=\mathbf{1}_2$. This vacuum
breaks SU(2)$\times$SU(2) spontaneously down to a diagonal SU(2), but
it keeps the latter SU(2) -- corresponding to isospin -- unbroken.

The target space of the Skyrme model, $\mathcal{M}\simeq$ SU(2)
$\simeq S^3$, has a nontrivial homotopy group 
\beq
\pi_3(\mathcal{M}) = \mathbb{Z}, 
\eeq
which admits solitons called Skyrmions. 
The topological degree $B \in \pi_3(S^3)$ is defined as
\beq
B = \frac{1}{2\pi^2} \int d^3x\; \mathcal{B}^0,
\label{eq:B}
\eeq
where the baryon charge density is given by
\beq
\mathcal{B}^0 &=& -\frac{1}{12}\epsilon^{ijk} \Tr[L_i L_j L_k].
\label{eq:Bcal}
\eeq
$B$ is often called the baryon number.

The model is a nonlinear sigma model, which means that a lot of
ambiguity is left in the potential.
The vacuum is at $U=\mathbf{1}_2$ around which small excitations of
the field correspond to physical pions.
Therefore one physical parameter that is known in the pion vacuum is
the pion mass, which is given by
\beq
m_\pi^2 = -2\left.\frac{\p V}{\p\Tr[U]}\right|_{U=\mathbf{1}_2}.
\eeq
Hence the traditional pion mass term is written as
\beq
V_1 = m_1^2 \left(1-\frac{1}{2}\Tr[U]\right),
\label{eq:V1}
\eeq
giving rise to a pion mass
\beq
m_\pi^2 = m_1^2.
\label{eq:pionmass}
\eeq

However, another potential, called the modified pion mass term is
given by
\cite{Marleau:1990nh,Piette:1997ce,Kudryavtsev:1999zm,Kopeliovich:2005vg} 
\beq
V_{02} = \frac{1}{2}m_{02}^2\left(1 - \frac{1}{4}\Tr[U]^2\right),
\label{eq:V2}
\eeq
which also yields Eq.~\eqref{eq:pionmass}, see also
Refs.~\cite{Davies:2009zza,Nitta:2012wi,Gudnason:2013qba,Gudnason:2014gla,Gudnason:2014hsa}.\footnote{This
  potential has two degenerate vacua allowing for a domain wall
  interpolating between them. } 
By just knowing the pion mass, we cannot distinguish between the
potentials $V_1$ and $V_{02}$ given in Eq.~\eqref{eq:V1} and
\eqref{eq:V2}, respectively.
The difference is that $V_{02}$ gives \emph{exactly} the pion mass
term, whereas $V_1$ gives the pion mass term as well as higher-order 
pion interactions, such as $(\pi^a\pi^a)^2$ and higher powers. 

In fact, from just the pion mass term, \emph{any} normalized linear
combination of the terms\footnote{See also
  Refs.~\cite{Marleau:1990nh,Davies:2009zza}.} 
\beq
V_{0n} = \frac{1}{n} m_{0n}^2\left(1 - \frac{1}{2^n}\Tr[U]^n\right),
\label{eq:V0n}
\eeq
gives rise to the physical pion mass around the vacuum
$U=\mathbf{1}_2$.

One aspect of this argument is that the pion mass is only the sum of
any of the terms $V_{0n}$ in Eq.~\eqref{eq:V0n}; the other side of the
same coin is that there is an enormous ambiguity in the nonlinearity
of the potential.

In particular, we can write a class of potentials
\beq
V_n = \frac{1}{n} m_n^2 \left(1 - \frac{1}{2}\Tr[U]\right)^n,
\label{eq:Vn}
\eeq
which for $n\geq 2$ gives no contribution to the pion mass in the
vacuum $U=\mathbf{1}_2$.\footnote{A recent paper considers this class
  of potentials in the BPS Skyrme model \cite{Ioannidou:2016kaj}. }

As we mentioned in the introduction, one of these potentials, namely
$V_4$ has received some attention recently, due to the fact that it
saturates a lower bound on the energy, giving a Skyrmion mass
proportional to the baryon number \cite{Harland:2013rxa}. 
Unfortunately, only the solution for $B=1$ (a single baryon) saturates
the energy bound \cite{Harland:2013rxa}.
However, solutions with baryon numbers larger than one have masses
quite close to the bound, yielding the possibility for relatively
small classical binding energies.
The model is therefore dubbed the lightly bound Skyrme model
\cite{Gillard:2015eia}. 

Let us contemplate for a moment what happens when adding a potential
$V_n$ of Eq.~\eqref{eq:Vn} to the Skyrme Lagrangian density. Since the
Skyrmion is a map from the target space $S^3$ to space
$\mathbb{R}^3\cup\{\infty\}\simeq S^3$, of positive degree, then at
least $B>0$ points in configuration space ($\mathbb{R}^3$) will attain
the value $U=-\mathbf{1}_2$, i.e.~the antipodal point to the vacuum on
the target space.
At these points, all the potentials $V_n$ (for any $n>0$) have their
maximum value. Since the map is topological, the Skyrmion cannot avoid
going over the points, but the effect is clear. The Skyrmion field
wants to get away from the antipodal points as quickly as possible,
but due to the presence of the kinetic term, this induces an effective 
repulsion between the antipodal points of the Skyrmion. The
implication is a reduction of the binding energy.
A similar effect was observed for the same potential in the baby
Skyrme model, where the authors called the baby Skyrmions aloof due to
the latter effect \cite{Salmi:2014hsa}.

\begin{figure}[!ht]
  \begin{center}
    \includegraphics[width=0.49\linewidth]{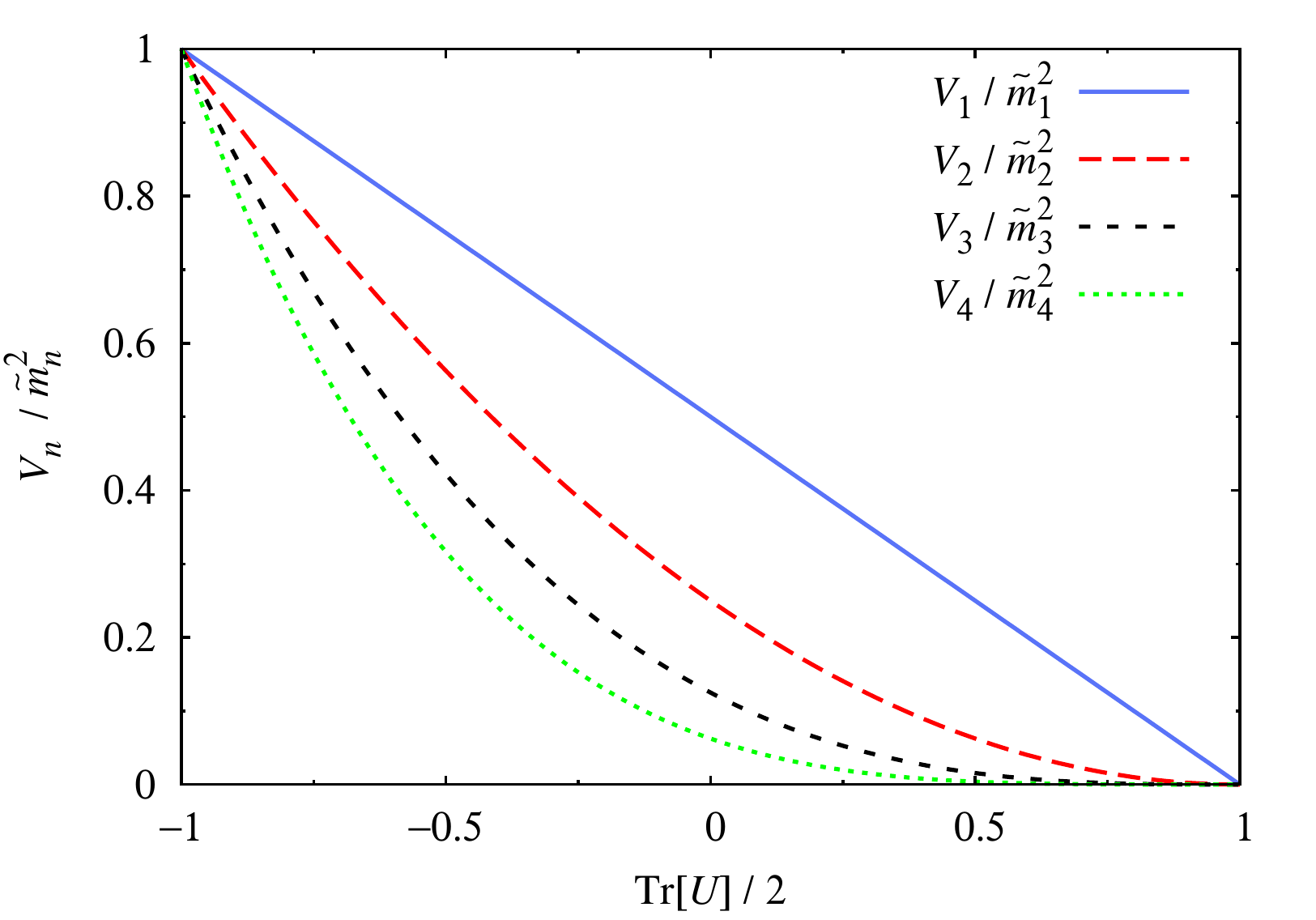}
  \end{center}
  \caption{Potentials $V_n/\tilde{m}_n^2$ normalized by the rescaled 
    masses as functions of $\Tr[U]/2$ for $n=1,2,3,4$. }
  \label{fig:pots}
\end{figure}

Let us define a rescaled mass
\beq
\tilde{m}_n\equiv \frac{2^{\frac{n}{2}} m_n}{\sqrt{n}}.
\eeq
At the antipodal point on the target space, $V_n/\tilde{m}_n^2$ tends
to unity. Therefore, if we now hold $\tilde{m}_n$ fixed and increase
$n$, nothing changes at the antipodal point, but the function goes to
zero faster the larger $n$ is.
It is now clear that the potential $V_n$ with larger $n$ induces
stronger repulsion than $V_n$ with a smaller $n$. In particular, the
repulsion is a monotonically increasing function of $n$.
Fig.~\ref{fig:pots} shows the potentials $V_n/\tilde{m}_n^2$ for
various values of $n$.

The potentials $V_n$ for $n>1$ are basically free parameters of the
theory as they are not directly measured (and are not related to the
pion mass). This is not the case for the potentials $V_{0n}$ whose sum
is constrained to be within reasonable range of the measured pion
mass.\footnote{The reason for not fixing the pion mass to the exact
  value measured in experiment is that the latter value is the pion
  mass in the pion vacuum, appropriate for describing pion
  physics. The pion mass relevant for the Skyrmion is the renormalized
  effective pion mass inside the baryon. This value is not necessarily
  the same, but is expected to be within a factor of a few within the
  measured value. }

As we mentioned in the introduction, the reduction of the binding
energy is of course more than welcome. However, the repulsion -- if
too excessive -- also leads to Skyrmions with different symmetries
than the platonic symmetries and in particular not preferring crystals
of alpha particles. Ref.~\cite{Gillard:2015eia} found that the
Skyrmion in the limit of large $m_4$ consists of $B$ spheres located
at the vertices of a face-centered-cubic (FCC) lattice.

In this paper, we will consider a more complicated potential
\beq
V = V_1 + V_2 + V_4,
\label{eq:pot}
\eeq
which depends on the parameters $m_1$, $m_2$ and $m_4$.
In light of the above discussion, it is clear that $V_4$ induces more 
repulsion than $V_2$ which in turn induces more repulsion than $V_1$. 
The value of $m_1$ is, however, not quite a free parameter; but $m_2$
and $m_4$ are.

Now let us consider the coefficients $c_2$ and $c_4$. The Skyrme
units correspond to $c_2=c_4=2$ where energies and lengths are given
in units of $f_\pi/(4e)$ and $2/(e f_\pi)$, respectively, see
Ref.~\cite{MantonSutcliffe}.
As the region where the repulsion is large, corresponding to smaller
binding energies, is where the parameters $m_2$ and $m_4$ are large,
we choose to use different values for the coefficients $c_2$ and
$c_4$, namely
\beq
c_2 = \frac{1}{4}, \qquad
c_4 = 1.
\eeq
Now the energies and lengths are given in units of $f_\pi/e$ and $1/(e
f_\pi)$, respectively.
When the normal Skyrme model units are used, a common choice of the
pion mass is $m_\pi=1$, which in our rescaled units corresponds to
$m_\pi=1/4$.

A mathematical problem is to find an energy bound for the Skyrme model
with the potential \eqref{eq:pot} and the closer the energies for
various $B$-Skyrmions are to the bound, the smaller the classical
binding energy must be.

Here, we are instead interested in a more difficult problem. We want
to get as close to the (best possible) energy bound as we can
\emph{and} at the same time keep the symmetries of the strongly bound
Skyrmions. In particular, we want the binding energy per nucleon of
$B=4$ to be larger than that of $B=5$ (and also that of $B=8$). This
latter condition implies that higher $B$ Skyrmions are composed by
crystals of alpha particles. 

This problem is of course somewhat difficult to address from a purely
mathematical angle. We therefore turn to numerical methods and
calculate numerical solutions in the next section.

\section{Numerical solutions}\label{sec:numerical}

In this section we embark on a large-scale numerical calculation of
many series of Skyrmion solutions in the parameter space spanned by
$\{m_1,m_2,m_4\}$ for $B=1,2,3,4,5$. We do not consider $B>5$ in this
paper due to the amount of computing resources needed for this
investigation. However, our analysis should be sufficient for having
only $B$ equal one through five.

Let us first mention the numerical method we will use to calculate the
numerical Skyrmion solutions. We will discretize space with the
finite-difference method using a fourth-order stencil and then cool
the partial differential equations (PDEs) with the relaxation method
until a static solution has been found to the accuracy that we
require. The relaxation method of course requires an initial condition
(configuration), for which we will use an appropriate rational map
Ansatz with the given baryon number $B$.
We will use the rational maps given in Ref.~\cite{Houghton:1997kg}. 

Let us define the observables that we calculate for each solution.
Of course the classical mass of the Skyrmion is an important
value. However, it will be convenient to evaluate the classical
binding energy 
\beq
\Delta_B = B E_1 - E_B,
\label{eq:EB}
\eeq
and in particular the \emph{relative} (classical) binding energy,
which we define as
\beq
\delta_B \equiv \frac{\Delta_B}{B E_1}
= 1 - \frac{E_B}{B E_1}.
\label{eq:deltaB}
\eeq
This observable is very easy to compare to experimental data as the
units drop out.
Comparing all solutions for all values of $B$, we define
\beq
\varepsilon_\delta(a) \equiv
\sum_{B}\left(\delta_B^{\rm exp} - \delta_B - a\right)^2,
\label{eq:epsilondelta}
\eeq
which measures the overall discrepancy between the solutions and the
experimental data for a given parameter-space point.
The parameter $a$ is introduced as an overall bias, reflecting the
fact that the ground state energy of the quantized 1-Skyrmion is the
classical mass plus spin-$\tfrac{1}{2}$ and isospin-$\tfrac{1}{2}$
contributions, whereas e.g.~the ground state energy of the 4-Skyrmion
is simply the classical energy.\footnote{The contribution from the
  spin and isospin quantization of the 1-Skyrmion to the energy
  modifies $\delta_B$ as 
  \beq
  \delta_B\to 1-\frac{E_B}{B(E_1+\epsilon_1)}
  = 1-\frac{E_B}{B E_1} + \frac{\epsilon_1 E_B}{B E_1^2}
  + \mathcal{O}(\epsilon_1^2),
  \eeq
  where we for simplicity use the parameter $a$ instead of the
  physical parameter $\epsilon_1$.
  There is also a contribution $\epsilon_B$, but it is typically a
  smaller effect. }
The ground states of the 2- and
3-Skyrmions are the spin-1, isospin-0 and spin-$\tfrac{1}{2}$,
isospin-$\tfrac{1}{2}$ states, respectively.
Nevertheless, the additional contribution to the ground state energy
for the 1-Skyrmion typically turns out to be larger than both that of 
the 2- and 3-Skyrmions.
This can be understood from the fact that the 2- and 3-Skyrmions are
larger resulting in larger moments of inertia and in turn smaller
quantum contribution to their energies. 

A more rigorous method would be to identify the symmetries of the
$B$-Skyrmions for each point in the parameter space and then quantize
their zero modes, incorporating the Finkelstein-Rubinstein constraints
for each of them, evaluating the moments of inertia tensors and
calculating their ground state energies.
For now, we will stick to just evaluating the classical binding
energies, knowing that they should be somewhat smaller than the
experimental values, but still in the ballpark. 

Another observable is the size of the Skyrmion, which we define in
terms of the baryon charge density \eqref{eq:Bcal} as
\beq
r_B^2 \equiv \frac{1}{2\pi^2 B}\int d^3x\; r^2 \mathcal{B}^0,
\eeq
where $r^2=x^2+y^2+z^2$ is a radial coordinate measured from the
center of the charge distribution.\footnote{A recent paper argues that
  using the baryon charge density for the volume/size is in some sense
  the natural way in Skyrme-like models \cite{Adam:2015zhc} (as
  opposed to using e.g.~the energy density). }
The length unit is just fitted to experimental data; therefore it will
prove convenient to use a \emph{relative} size
\beq
\rho_B \equiv \frac{r_B}{r_1},
\eeq
where $r_B=\sqrt{r_B^2}$ and $\rho_B$ is given in units of the size of
the $B=1$ solution.
Comparing again all solutions for all values of $B$, we define
\beq
\varepsilon_\rho \equiv
\sum_{B}\varepsilon_{\rho_B}, \qquad
\varepsilon_{\rho_B} \equiv \rho_B^{\rm exp} - \rho_B.
\label{eq:epsilonrho}
\eeq
Notice that we do not square the summands so that the sign will be
evident (negative if the solutions are too large and positive if
not).\footnote{Although this definition allows for the caveat that
  some cancellation between different $B$-Skyrmion sizes takes place,
  this will not be an issue as all the $B$-Skyrmions are generally too
  small compared to nuclei. }

Finally, an observable which gives a good handle on the accuracy, is
the numerically integrated baryon number \eqref{eq:B}. Our solutions 
will be equal to the integer $B$ with an accuracy in the range of
$[0.16\%,0.019\%]$ (with an overall average around $0.052\%$).

For the $B=1$ sector, we calculate all the solutions with very high
accuracy using the ordinary differential equation (ODE) derived from
the Lagrangian density \eqref{eq:L} with the hedgehog Ansatz:
$U=\mathbf{1}_2\cos f(r)+i\tau\cdot\hat{x}\sin f(r)$. The ODE reads 
\begin{align}
  c_2\left(f_{rr} + \frac{2}{r}f_r - \frac{\sin 2f}{r^2}\right)
  +c_4\left(\frac{2\sin^2(f)f_{rr}}{r^2}
  + \frac{\sin(2f)f_r^2}{r^2}
  - \frac{\sin(2f)\sin^2f}{r^4}\right) \non
  = m_1^2\sin f + m_2^2(1-\cos f)\sin f + m_4^2 (1-\cos f)^3 \sin f,
\end{align}
where $f_r\equiv \p_r f$, etc.
The solution of the above equation yields
$E_1(m_1,m_2,m_4)$ with very high accuracy (better than the $10^{-6}$
level).
Let us now comment on how we calculate the energy for the $B>1$
solutions. As the $B=1$ sector is very accurate, we need a precise
estimate of the energy for the higher $B$ solutions in order to
calculate the classical binding energy \eqref{eq:EB} and in turn the
relative classical binding energy \eqref{eq:deltaB} (otherwise we will
underestimate them).
First, we find our solution relaxed down to the accuracy level such
that all equations of motion are satisfied better than the $10^{-3}$
level locally. From this point on, the energy as function of
relaxation time, $\tau$ (steps), is then fitted to an exponential 
curve and this process is continued until the accuracy of the
exponential fit has converged to a given accuracy.
Then we take the $\tau\to\infty$ limit of the exponential as an
estimate of the asymptotic energy value. This trick is very precise
and saves some computation time. Now, since our finite-difference
lattice is also just an approximation to the continuous field and the 
fact that the Skyrmion charge is a convex function (resulting in
$B_{\rm numerical}<B$), we compensate the final result by
$B/B_{\rm numerical}$.
The final result has the form
\beq
E_B \simeq \frac{B}{B_{\rm numerical}}\times
\frac{E_{B,{\rm numerical}}(\tau_0)E_{B,{\rm numerical}}(\tau_2)
  - E_{B,{\rm numerical}}^2(\tau_1)}{E_{B,{\rm numerical}}(\tau_0)
  - 2E_{B,{\rm numerical}}(\tau_1) + E_{B,{\rm numerical}}(\tau_2)},
\eeq
where $\tau_0$ is the relaxation time where the solution is good
enough for the initial accuracy level (EOMs at the $10^{-3}$ level,
locally), $\tau_2$ is the final relaxation time where the exponential
fit is precise enough and $\tau_1 = (\tau_0 + \tau_2)/2$.
After this complicated process of estimating the energy of the
Skyrmion solution, we check for the $B=1$ sector that we obtain the
energies within an accuracy of about $2.7\times 10^{-4}$ or better. 

We are now ready to present the results in the next subsections,
for vanishing and non-vanishing pion mass, respectively, and finally
the effect of semi-classical zero-modes quantization.

\subsection{Zero pion mass}

We will begin with taking a vanishing pion mass $m_1=0$ and scan (a
part of) the $(m_2,m_4)$ parameter space. In the next subsection we
will consider the inclusion of the pion mass.

\begin{figure}[!htp]
  \begin{center}
    \includegraphics[width=0.49\linewidth]{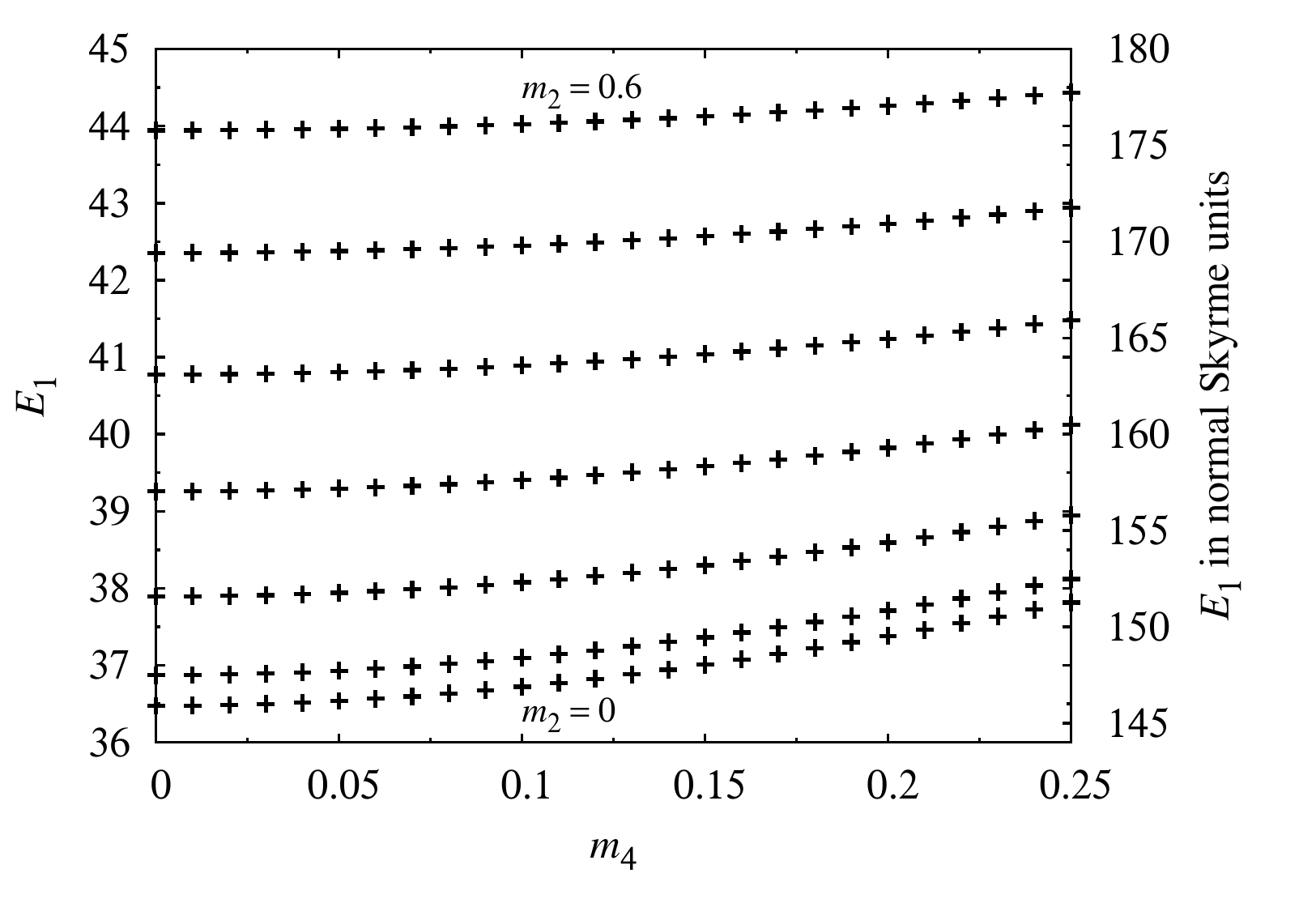}
  \end{center}
  \caption{Energy of the $B=1$ Skyrmion with various values in the
    $(m_2,m_4)$-parameter space. The series of points is for
    $m_2=0,0.1,0.2,0.3,0.4,0.5,0.6$ with $m_2$ increasing from bottom
    to top. The left-hand scale shows the units we are using in this
    paper while the right-hand scale shows the normal Skyrme units. }
  \label{fig:e1}
\end{figure}

We start by calculating the Skyrmion energies in the $B=1$ sector, for
which as we mentioned above use simply the ODE. This is very precise
and we will use these energies as the basis to calculate the binding
energies for the higher-$B$ Skyrmion solutions.
Fig.~\ref{fig:e1} shows the energies in our units (which are
normalized differently than the normal Skyrme units) for solutions in
the $(m_2,m_4)$-parameter space. 
For comparison Fig.~\ref{fig:e1} has the normal Skyrme units on the
right-hand scale.
Throughout this section the ranges of the masses in the parameter
space will be chosen as $m_4$ from 0 to 0.25 with steps of 0.01 and
for $m_2$ from 0 to 0.6 with steps of 0.1. 

Now we are ready to calculate the higher-$B$ Skyrmions. We use very
small increasing/decreasing steps for $m_4$ and use the latest data
point as an initial condition for the next one. We tried going both
from the $(m_2,m_4)=(0,0)$ point and upwards in masses and the reverse
in order to check that the solutions found are really the minimizers
of the energy for the given value of $(m_2,m_4)$.
As we mentioned already, the $(m_2,m_4)=(0,0)$ point is calculated
with the initial conditions constructed from the rational map
Ans\"atze of Ref.~\cite{Houghton:1997kg}.
If the steps in, for instance $m_4$, are too large then the direction
(increasing or decreasing of the mass) may give different solutions to
the approximated accuracy levels chosen for numerical calculations.
Therefore we use quite small steps and check that the results do not
change much by reversing the direction (we found some critical points
in parameter space where the solutions did shift a bit, but it will
not have essential consequences for our purpose here). 
Figs.~\ref{fig:f2} through \ref{fig:f5} show isosurfaces of the baryon
charge density at half maximum values for the chosen part of parameter
space in the $(m_2,m_4)$-plane (only every second solution in the
$m_4$-direction is shown in these figures due to space limitations). 
The coloring adapted here is chosen such that the pions are normalized
to a unit vector $(\hat{\pi}^2=1)$ and $\hat{\pi}_1$ determines the
lightness whereas $\hat{\pi}_3+i\hat{\pi}_2$ is mapped to the hue of
the color circle (the coloring scheme is similar to that adapted in
Refs.~\cite{Manton:2011mi,Feist:2012ps}, see also
Ref.~\cite{Gudnason:2014jga}).

\begin{figure}[!tp]
  \begin{center}
    \mbox{
      \includegraphics[width=0.49\linewidth]{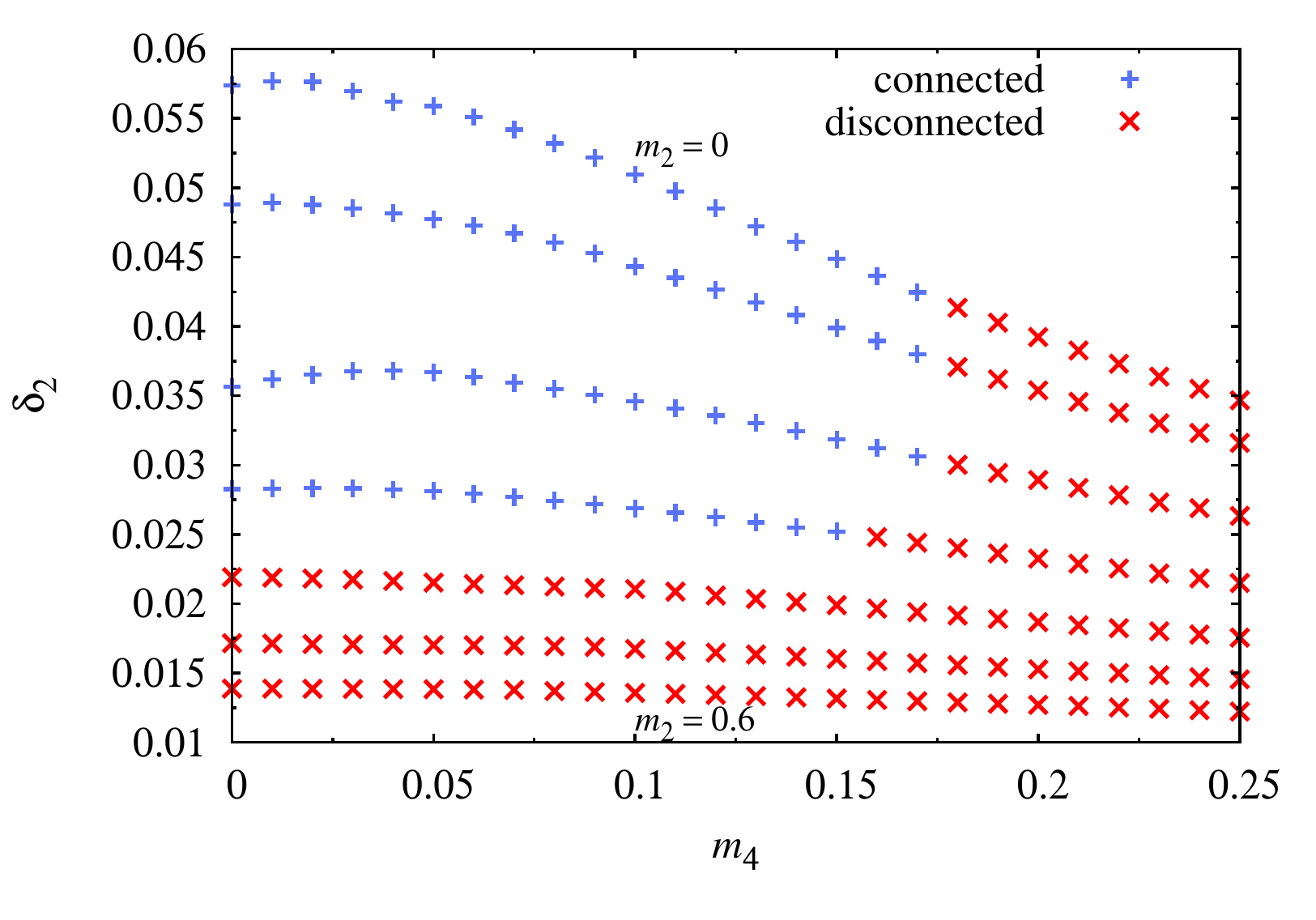}
      \includegraphics[width=0.49\linewidth]{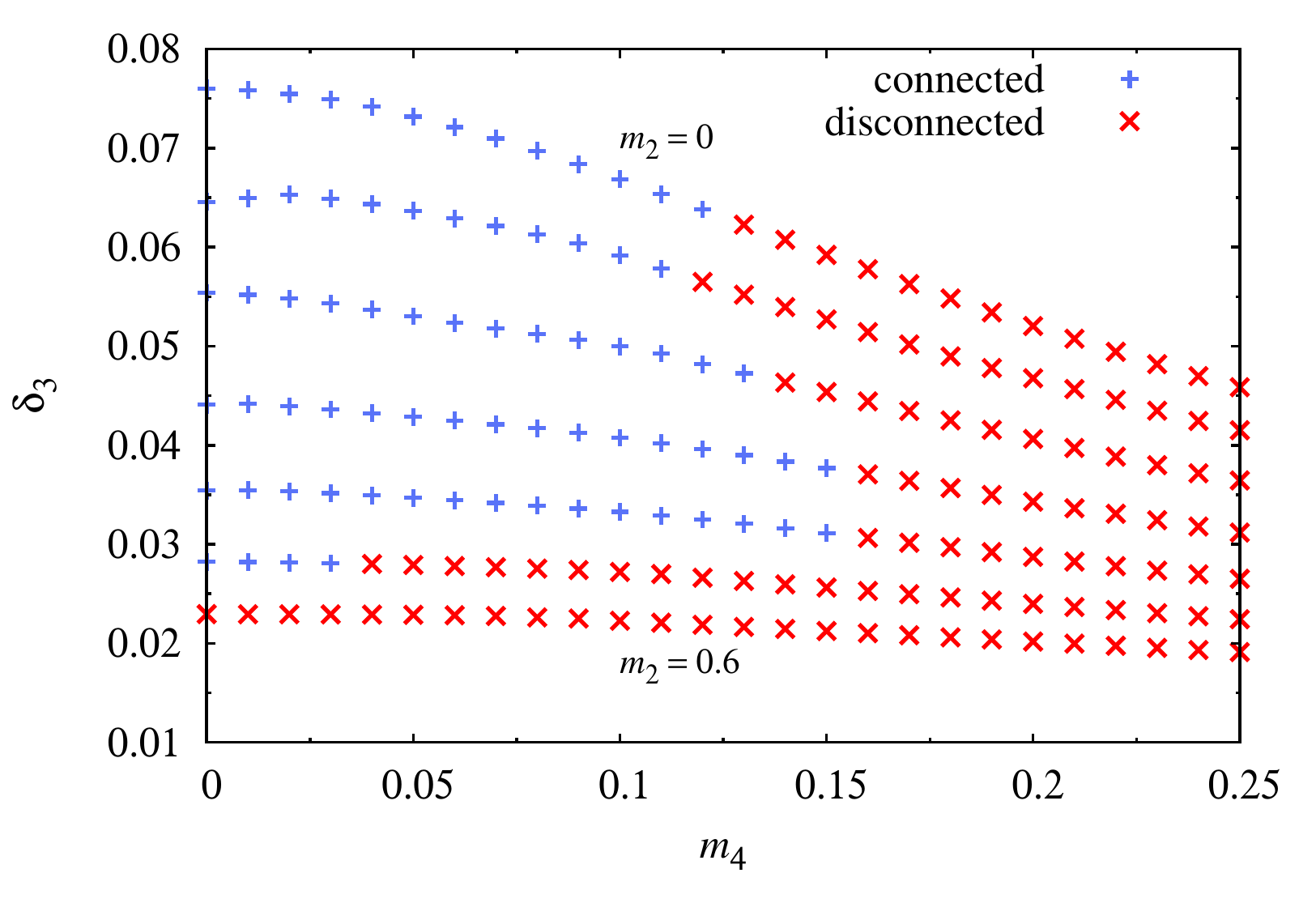}}
    \mbox{
      \includegraphics[width=0.49\linewidth]{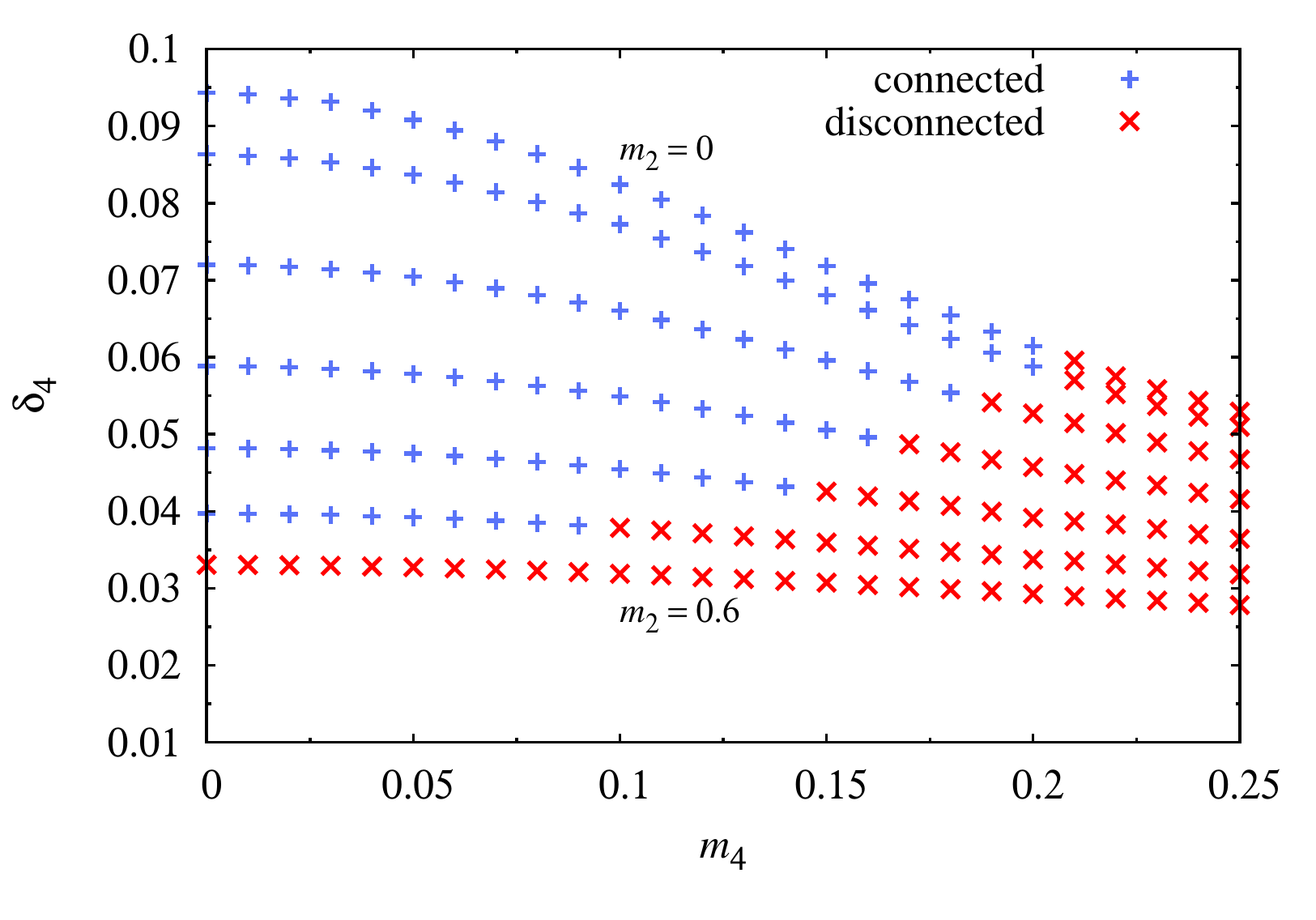}
      \includegraphics[width=0.49\linewidth]{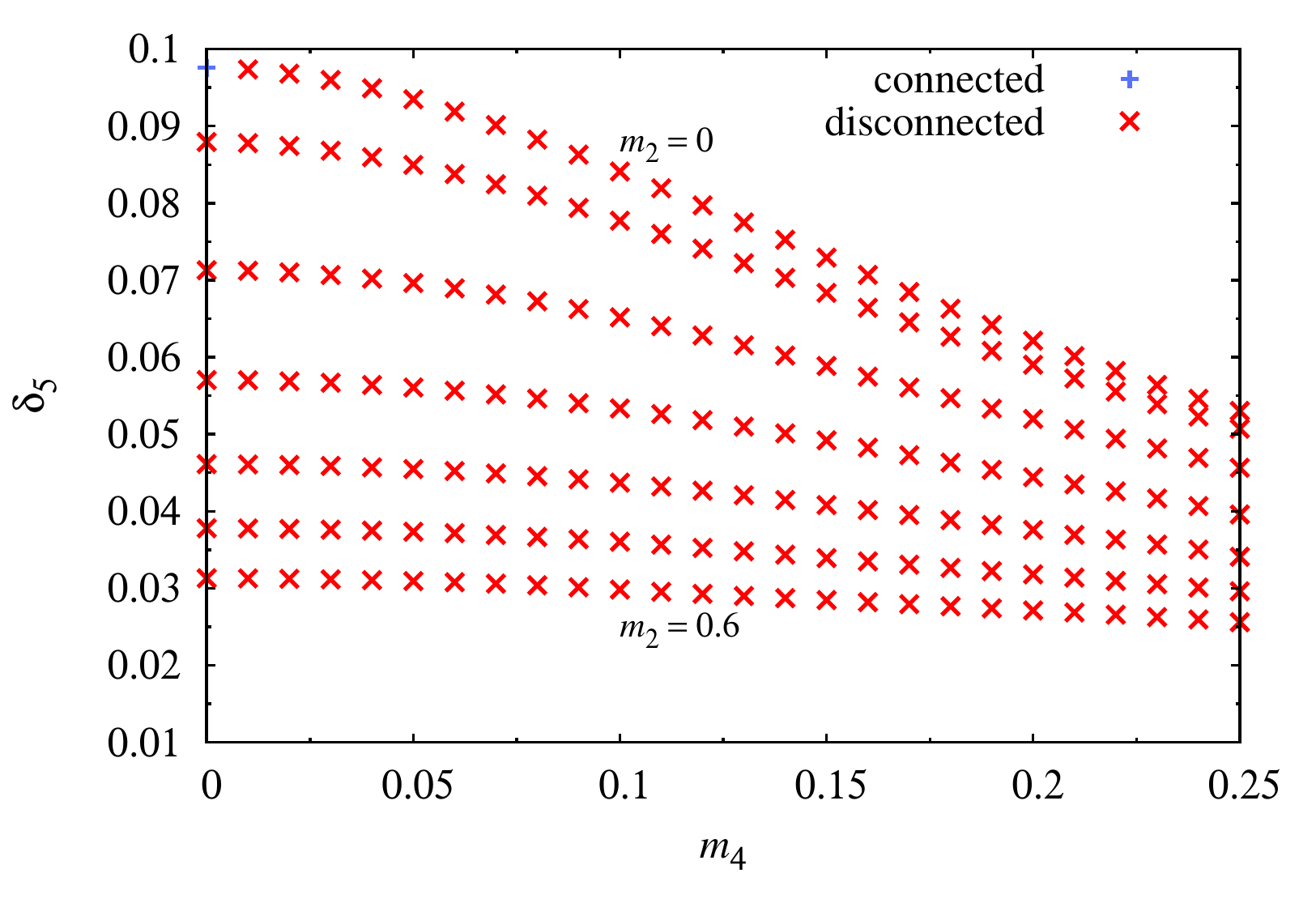}}
  \end{center}
  \caption{Relative classical binding energies $\delta_B$ for
    $B=2,3,4,5$. The series of points is for
    $m_2=0,0.1,0.2,0.3,0.4,0.5,0.6$ with $m_2$ increasing from top to
    bottom. The blue crosses ($+$) are connected isosurfaces at
    half-maximum baryon charge density while the red $x$s ($\times$)
    are disconnected.  } 
  \label{fig:rbe}
\end{figure}

Now that we have the data for a bunch of Skyrmion solutions, we begin
by calculating the classical binding energies for the different
points. 
Fig.~\ref{fig:rbe} shows the relative classical binding energies for
all the solutions and the blue crosses represent connected Skyrmions
(for the baryon charge density at half-maximum values), whereas the
red $x$s are disconnected.
Of course it is a bit arbitrary to choose connectedness at half the
maximum value of the baryon charge density; any other reasonable value
may be just as good and shift the connected/disconnected lines of the
figures.
Nevertheless, it is clear that in the far blue area the platonic
symmetries are still unbroken, whereas in the far red area the
Skyrmions are spheres at the vertices of an FCC lattice.

What we seek is to find a region in parameter space where the binding
energy is decreased with respect to that of the normal Skyrme model
and where the platonic symmetries are more or less still present. At
least the symmetries of the $B=4$ cubic Skyrmion would be preferable
to maintain, as it provides a number of phenomenologically appealing
properties as we mentioned in the introduction. 

The lesson we learn from all these data points is that increasing
$m_4$ (from zero) does indeed lower the binding energy as
expected. However, long before the binding energies of experimental
data are reached, the symmetries of the Skyrmions change from platonic
symmetries to the FCC lattice.
On the other hand, increasing $m_2$ (again from zero) has the same
qualitative effect; namely it decreases the binding energy and
eventually breaks the platonic symmetries to the same FCC lattice
structure of aloof Skyrmions.
The difference, however, is that the binding energies obtained before
the symmetries change are far lower when using $V_2$ than when using
$V_4$. Consider the $B=4$ sector in Fig.~\ref{fig:rbe}. If we regard
the boundary between the blue and red dots as some sort of measure
of change of symmetry, then the $m_2=0$ branch reaches classical
binding energies of about 6\%, whereas the $m_4=0$ branch goes down
below 4\%. 

Moreover, it is observed from Fig.~\ref{fig:rbe} that when
$m_2=0$, the binding energy does go down when increasing
$m_4$. However, when $m_2$ is large, increasing $m_4$ does not lower
the binding energy substantially before it breaks the platonic
symmetries down to the FCC lattice symmetries.
Therefore, if we insist on keeping the old symmetries of the normal
Skyrme model, then we can basically turn off the potential $V_4$ and
work with just $V_2$. If however we prefer the FCC lattice symmetries,
then $V_4$ is a suitable potential that lowers the classical binding
energies, but so is $V_2$.

\begin{figure}[!tp]
  \begin{center}
    \mbox{
      \includegraphics[width=0.49\linewidth]{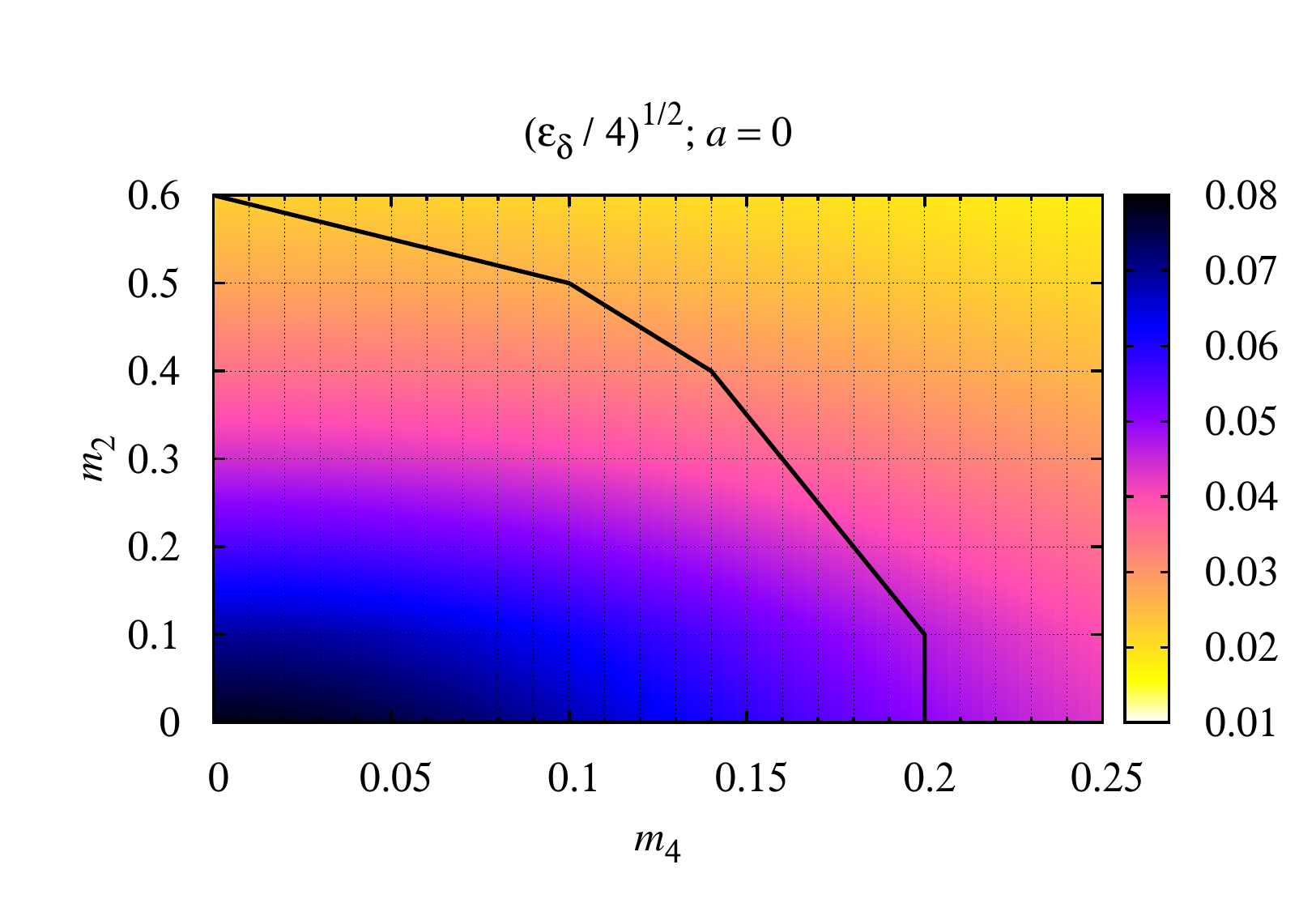}
      \includegraphics[width=0.49\linewidth]{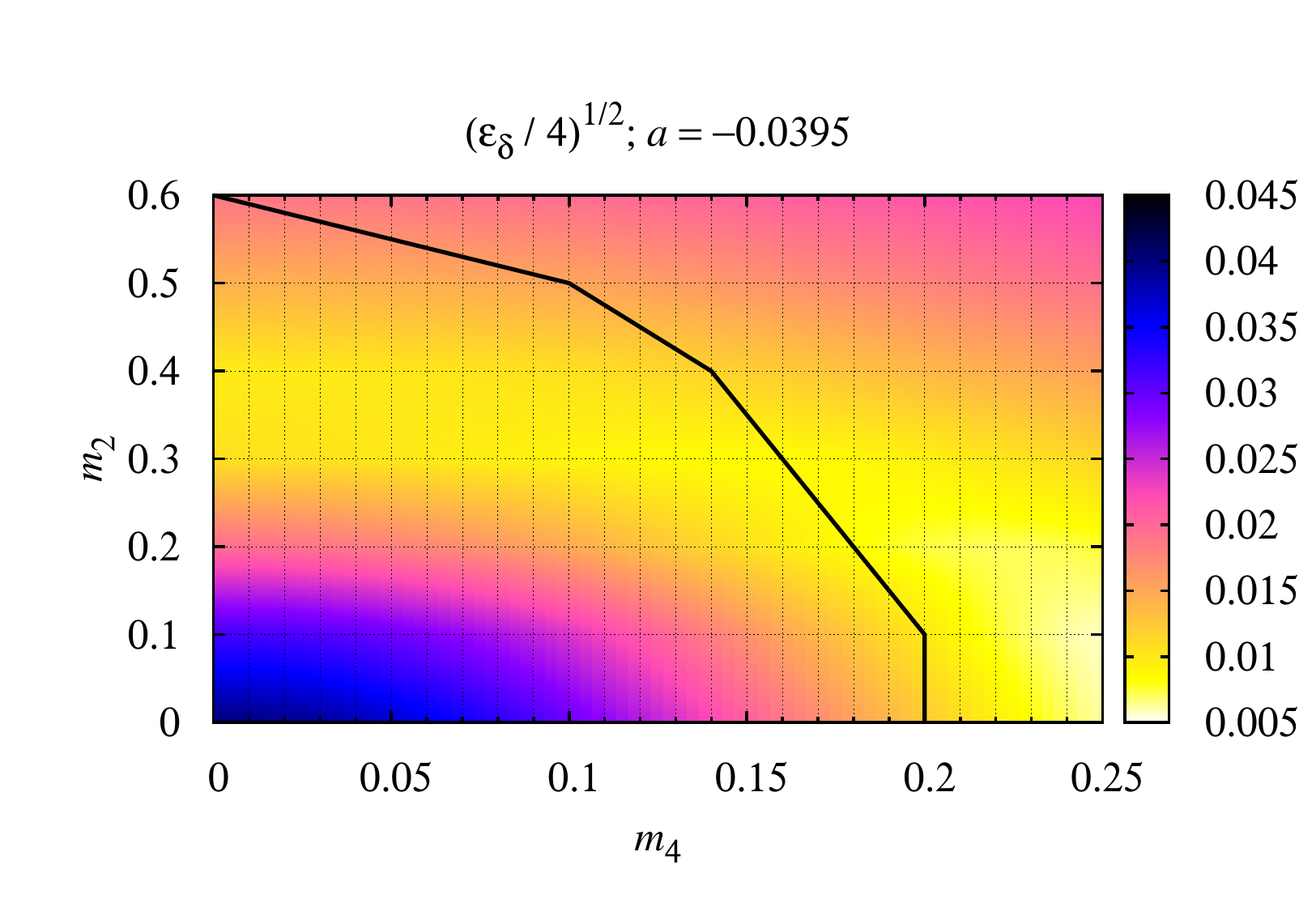}}
  \end{center}
  \caption{Fits of the relative binding energies $\delta_B$ summed
    up in the function $\varepsilon_\delta$.
    $\sqrt{\varepsilon_\delta/4}$ corresponds to the average
    discrepancy of the classical binding energy, which ranges from
    about 8\% to 1\%. 
    The black line shows where the $B=4$ Skyrmion splits up into
    disconnected pieces at the level of the isosurfaces at the
    half-maximum value of the baryon charge density. 
    The zero point, $a$ is fitted in the right panel of the figure,
    which corresponds to ignoring the $B=1$ Skyrmion's energy (that
    is, fitting just the shape of the remaining binding energies). } 
  \label{fig:errdelta}
\end{figure}

\begin{table}[!ht]
  \begin{center}
    \caption{Experimental values for nuclear masses. }
    \label{tab:nuclearbinding}
    \begin{tabular}{ll}
      ${}^1$H  & 1.007825\\
      ${}^2$H  & 2.014101\\
      ${}^3$He & 3.016029\\
      ${}^4$He & 4.002603\\
      ${}^5$He & 5.012057
    \end{tabular}
  \end{center}
\end{table}

Considering now the function \eqref{eq:epsilondelta}. This function is 
a least-squares fit function of the parameter space to
experimental data for the nuclear binding energies.
We use the experimental values shown in
Tab.~\ref{tab:nuclearbinding}. 
Fig.~\ref{fig:errdelta} shows the fit in the calculated part of
parameter space. The black line shows where the $B=4$ Skyrmion splits 
up into disconnected pieces at the level of the isosurfaces at the
half-maximum value of the baryon charge density. 
The left panel of the figure is the real fit of the classical binding
energies to the experimental data, whereas the right panel shows a fit
where the value $a$ has been optimized to improve the fit (shape fit
only). 
The physical meaning is that if the energy of the $B=1$
Skyrmion is reduced about 4\%, then the preferred region of the fit is
within the boundary of the black line and thus the platonic symmetries
remain while the classical binding energies of the higher-$B$
Skyrmions match reasonably well the experimental values.
Had the best value for $a$ been a positive value, then semi-classical
quantization could be a fix to this problem; but since it is negative
then quantization will only exacerbate the problem.

\begin{figure}[!tp]
  \begin{center}
    \mbox{
      \includegraphics[width=0.49\linewidth]{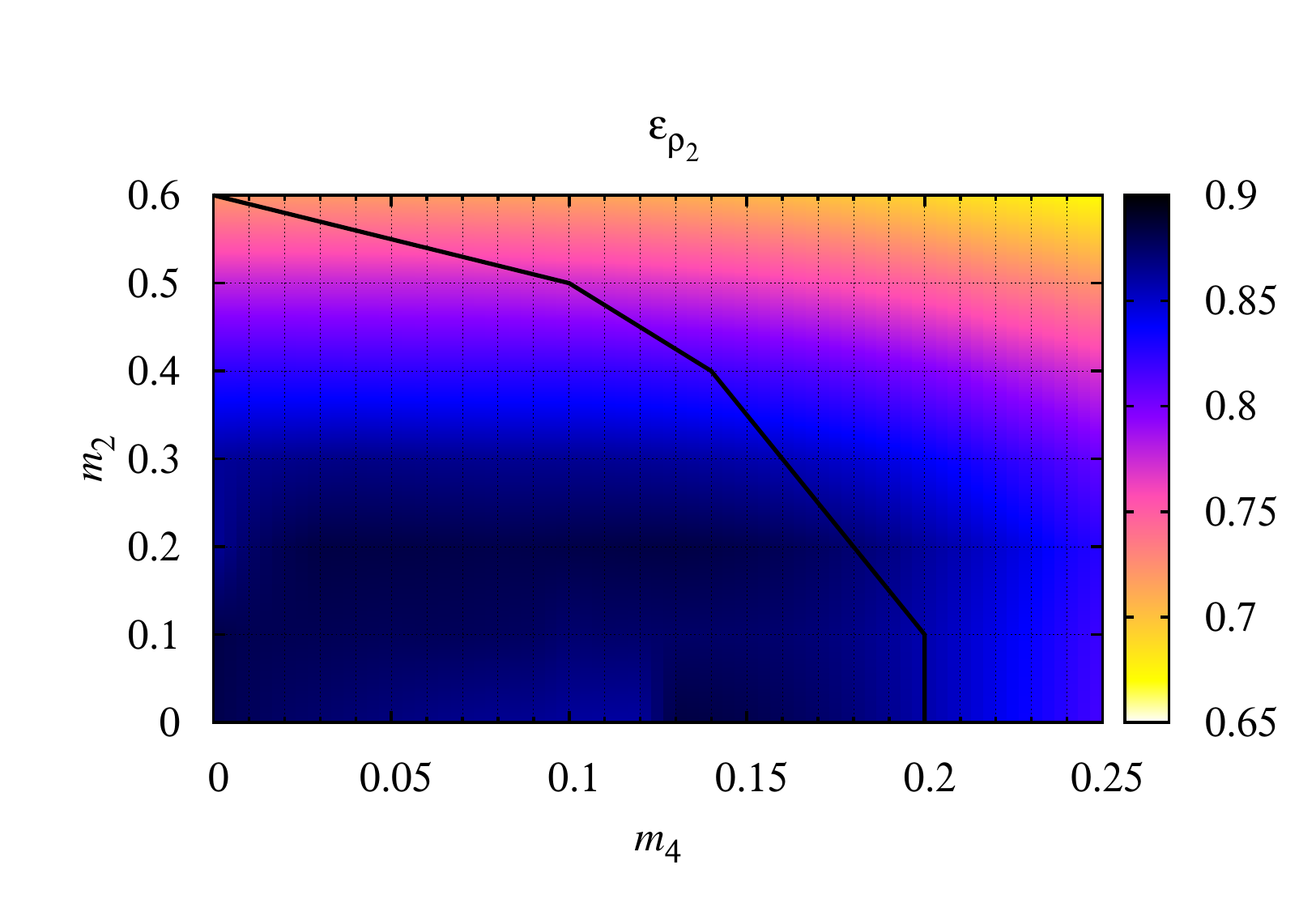}
      \includegraphics[width=0.49\linewidth]{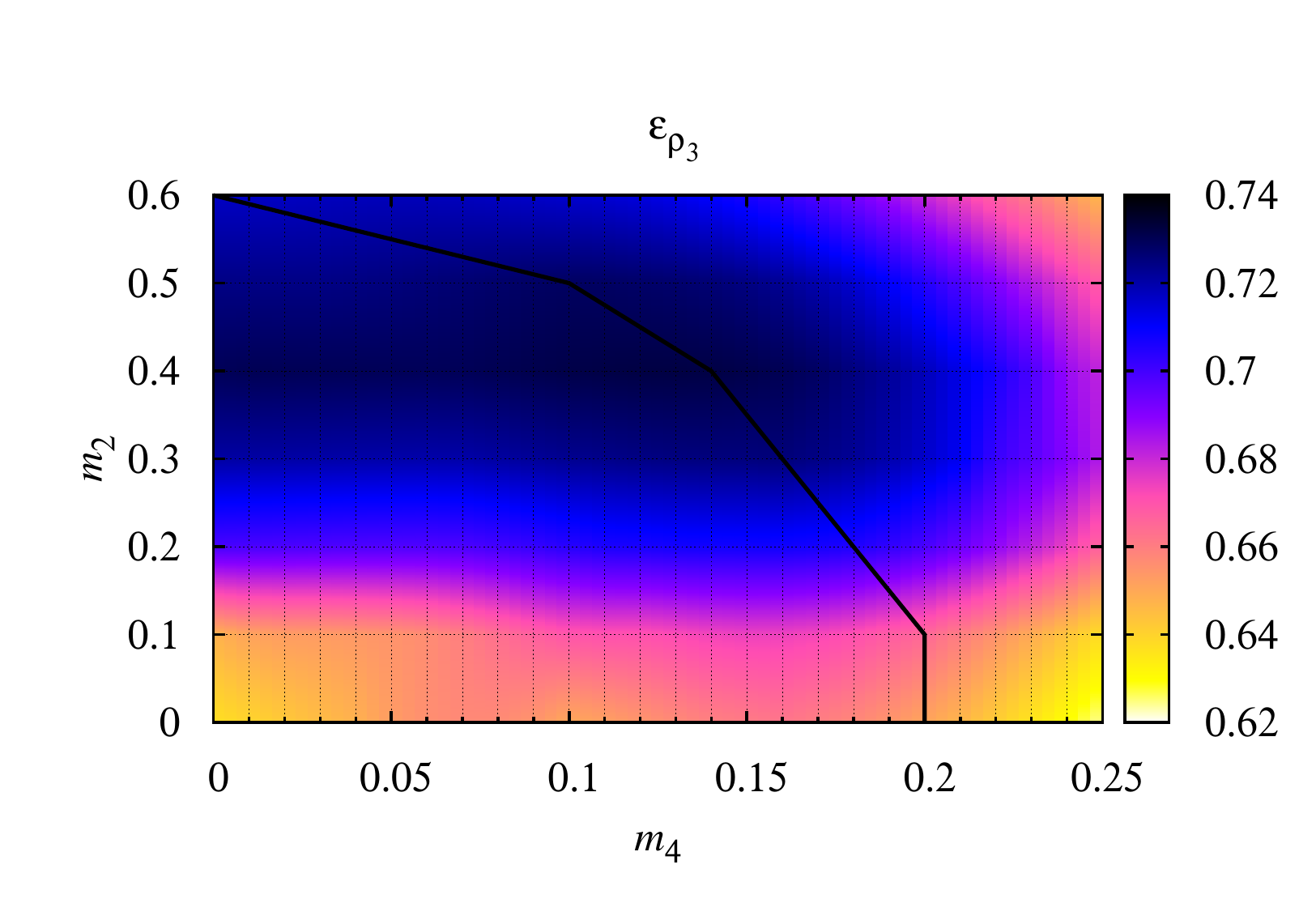}}
    \mbox{
      \includegraphics[width=0.49\linewidth]{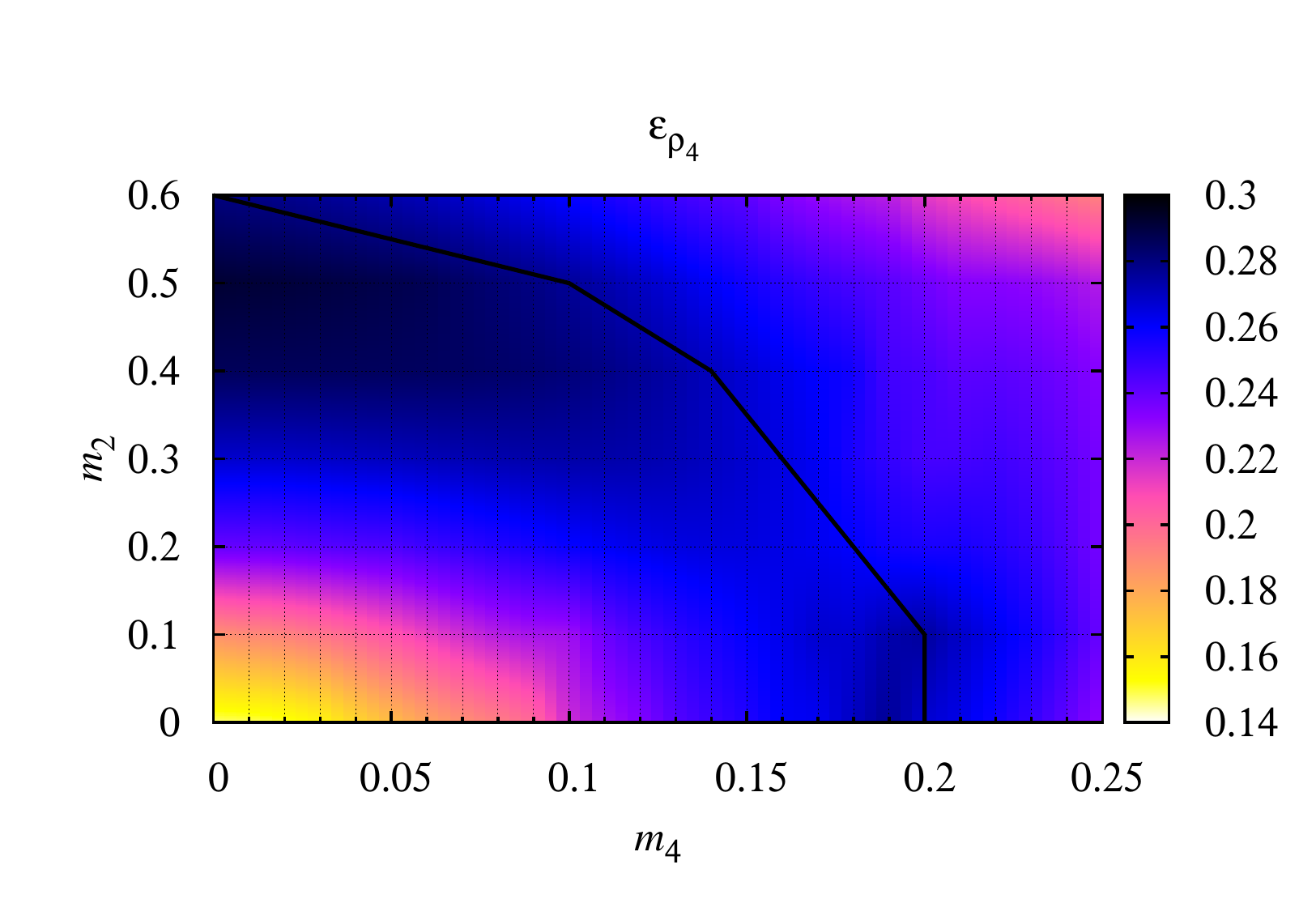}
      \includegraphics[width=0.49\linewidth]{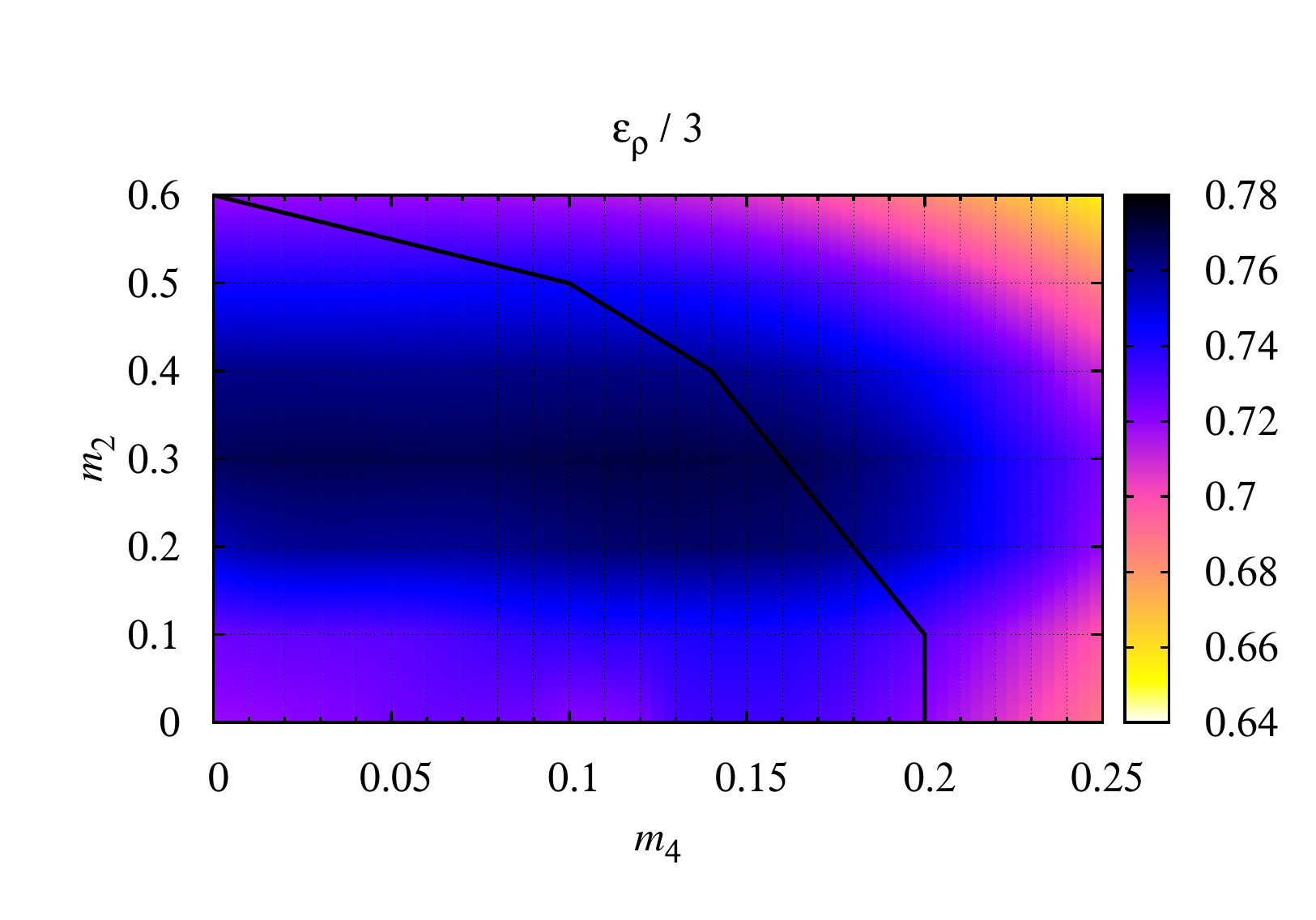}}
  \end{center}
  \caption{Fits of relative sizes, separately for $B=2,3,4$ and at
  last the mean fit of the same three Skyrmion sectors. Positive
  values indicate that the Skyrmion size is too small compared with
  experimental value for the nucleus (see
  Eq.~\eqref{eq:epsilonrho}). 
  The value of $\varepsilon_\rho$ corresponds roughly to the relative
  mismatch with data, which is in the range of 14\% to 90\%.
  The black line shows again where the $B=4$ Skyrmion splits up into
  disconnected pieces at the level of the isosurfaces at the
  half-maximum value of the baryon charge density. } 
  \label{fig:errrho}
\end{figure}

\begin{table}[!ht]
  \begin{center}
    \caption{Experimental values for charge radii
      \cite{Angeli201369}. } 
    \label{tab:chargeradii}
    \begin{tabular}{lc}
      ${}^1$H  & 0.8783 fm\\
      ${}^2$H  & 2.1421 fm\\
      ${}^3$He & 1.9661 fm\\
      ${}^4$He & 1.6755 fm\\
      ${}^5$He & --
    \end{tabular}
  \end{center}
\end{table}

Next we will consider a rough fit of the sizes of the Skyrmion
solutions to the experimental values of charge radii of nuclei.
The experimental values used here are shown in
Tab.~\ref{tab:chargeradii}.
Of course the charge radius is not quite the size of the nucleus, but
we take that as a good approximation to the latter.
Fig.~\ref{fig:errrho} shows the fits of the Skyrmion sizes to the
experimental data for the $B=2,3,4$ sectors as well as the average
fit of all three sectors.

The qualitative information that can be read off of
Fig.~\ref{fig:errrho} is that the 2-Skyrmion and the 3-Skyrmion are
generally too small. The 4-Skyrmion has about the right size when the
potentials are turned off, but then the binding energies are too
large.
It is interesting to note that the Skyrmion size is increased by the
addition of the sixth-order potential, which is the backbone of the
BPS Skyrme model \cite{Adam:2010fg,Adam:2010ds}, see also
Ref.~\cite{Gudnason:2014jga}.

\subsection{Nonzero pion mass}

Now we consider a physical value of the pion mass, which corresponds
to $m_1=1/4$ (this is equal to $m_\pi=1$ in the normal Skyrme
units). This value is commonly used in Skyrmion calculations, but
other values could also be considered.
Here we are mostly interested in the qualitative effect on our results 
with the addition of the pion mass.

As the common lore is that for $B\leq 7$ the qualitative effect of
the addition of the pion mass is rather small, we would a priori not
expect big changes with respect to the last subsection.
However, as we will see shortly, some changes do occur. 

\begin{figure}[!tp]
  \begin{center}
    \mbox{
      \includegraphics[width=0.49\linewidth]{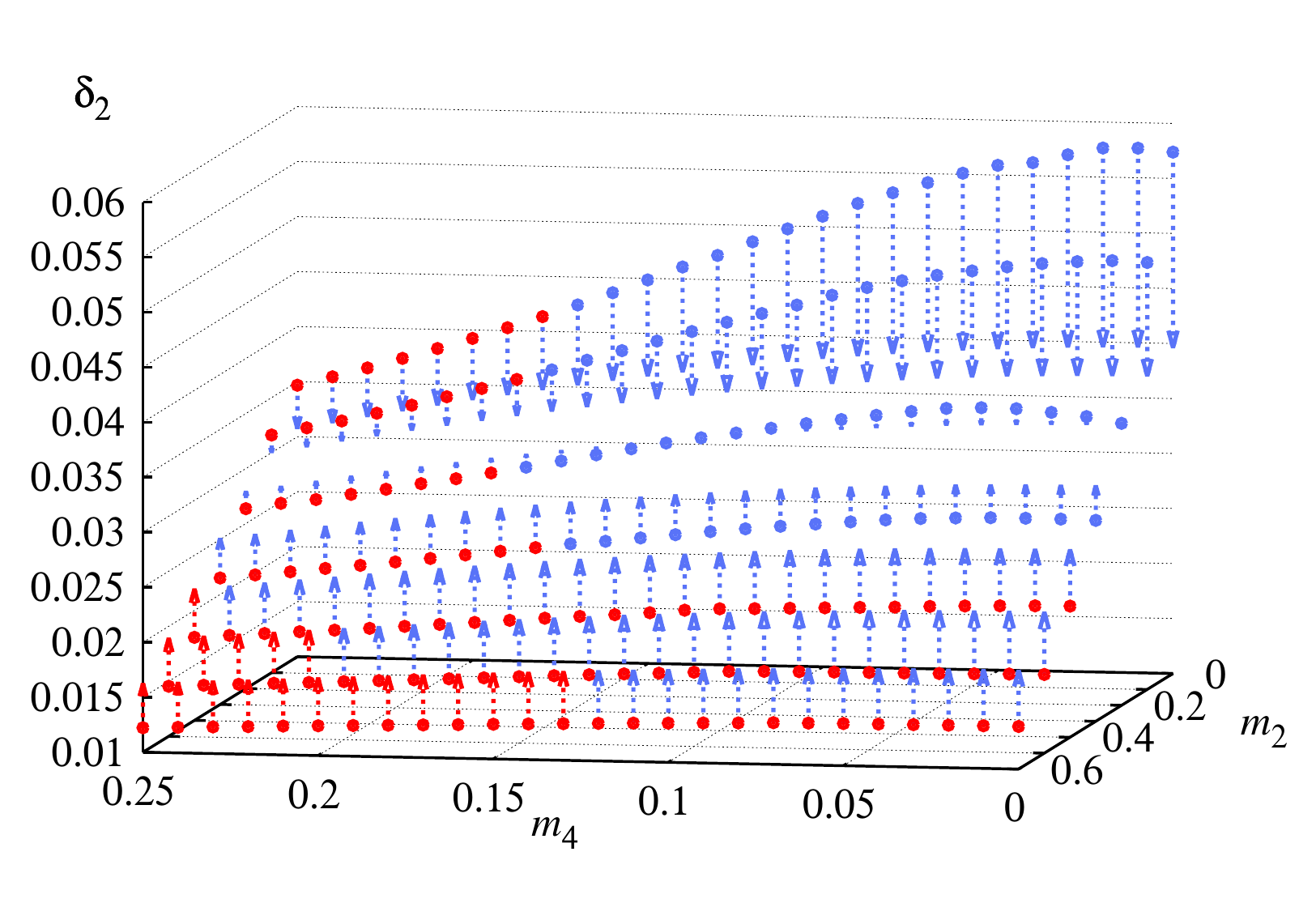}
      \includegraphics[width=0.49\linewidth]{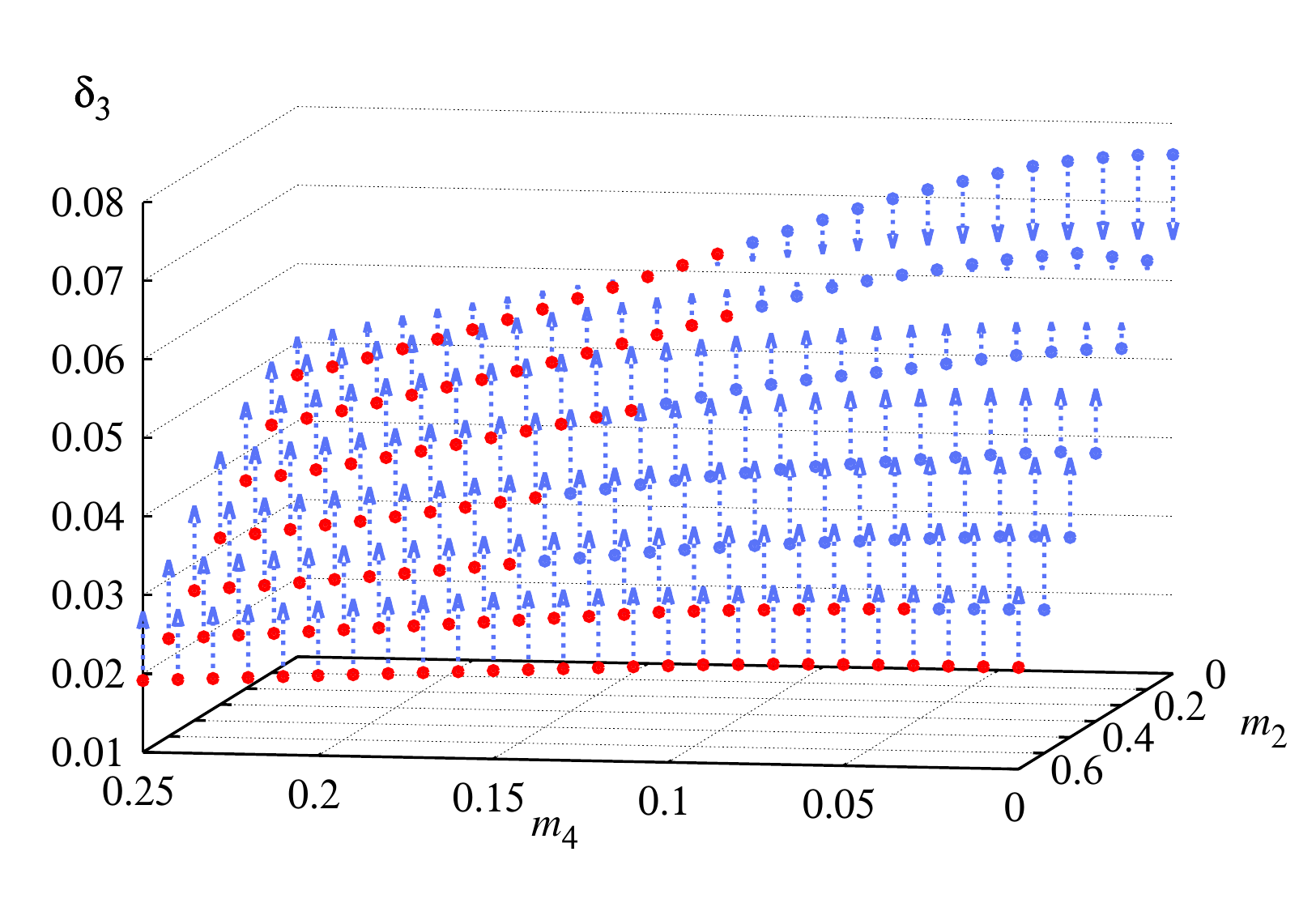}}
    \mbox{
      \includegraphics[width=0.49\linewidth]{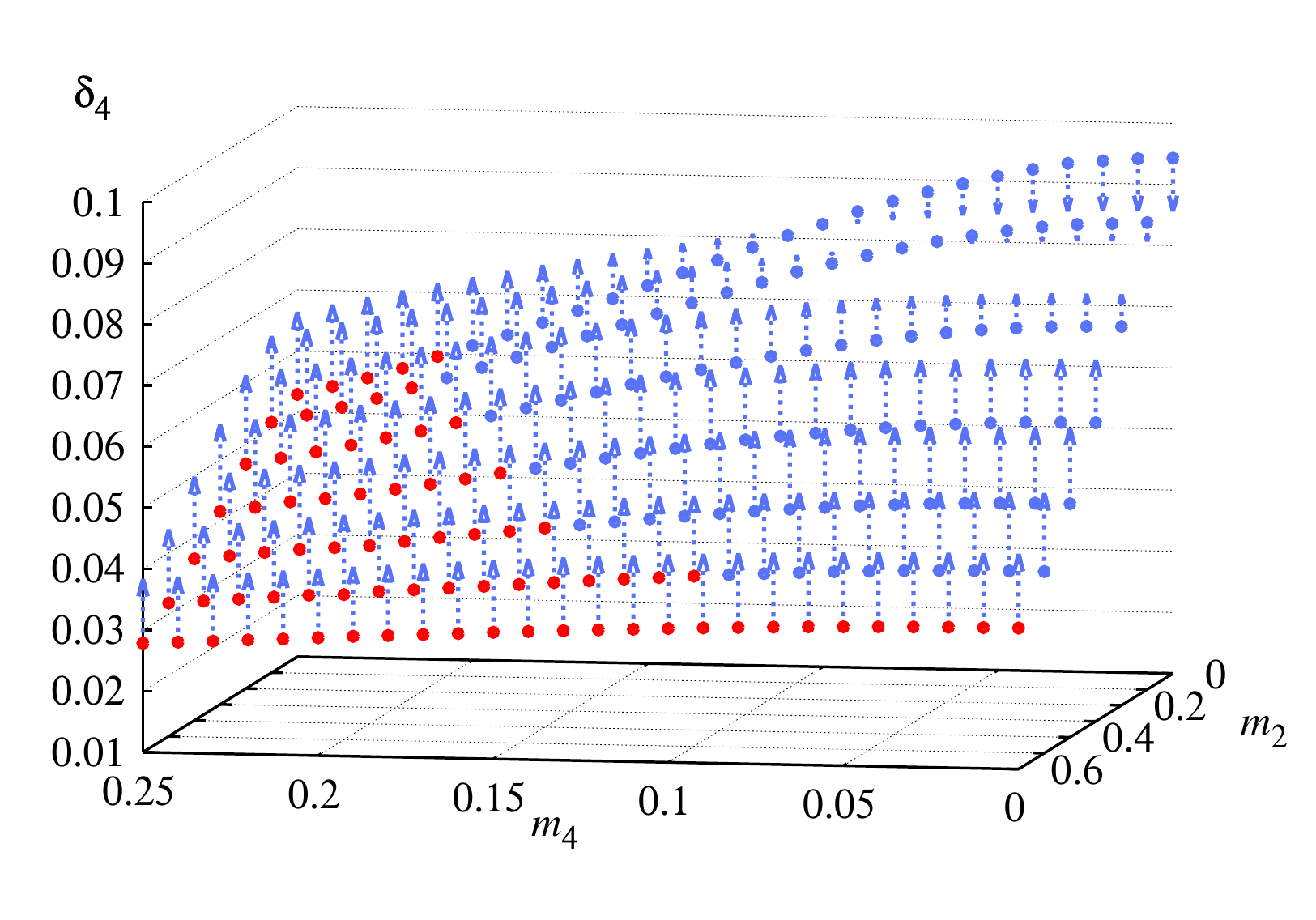}
      \includegraphics[width=0.49\linewidth]{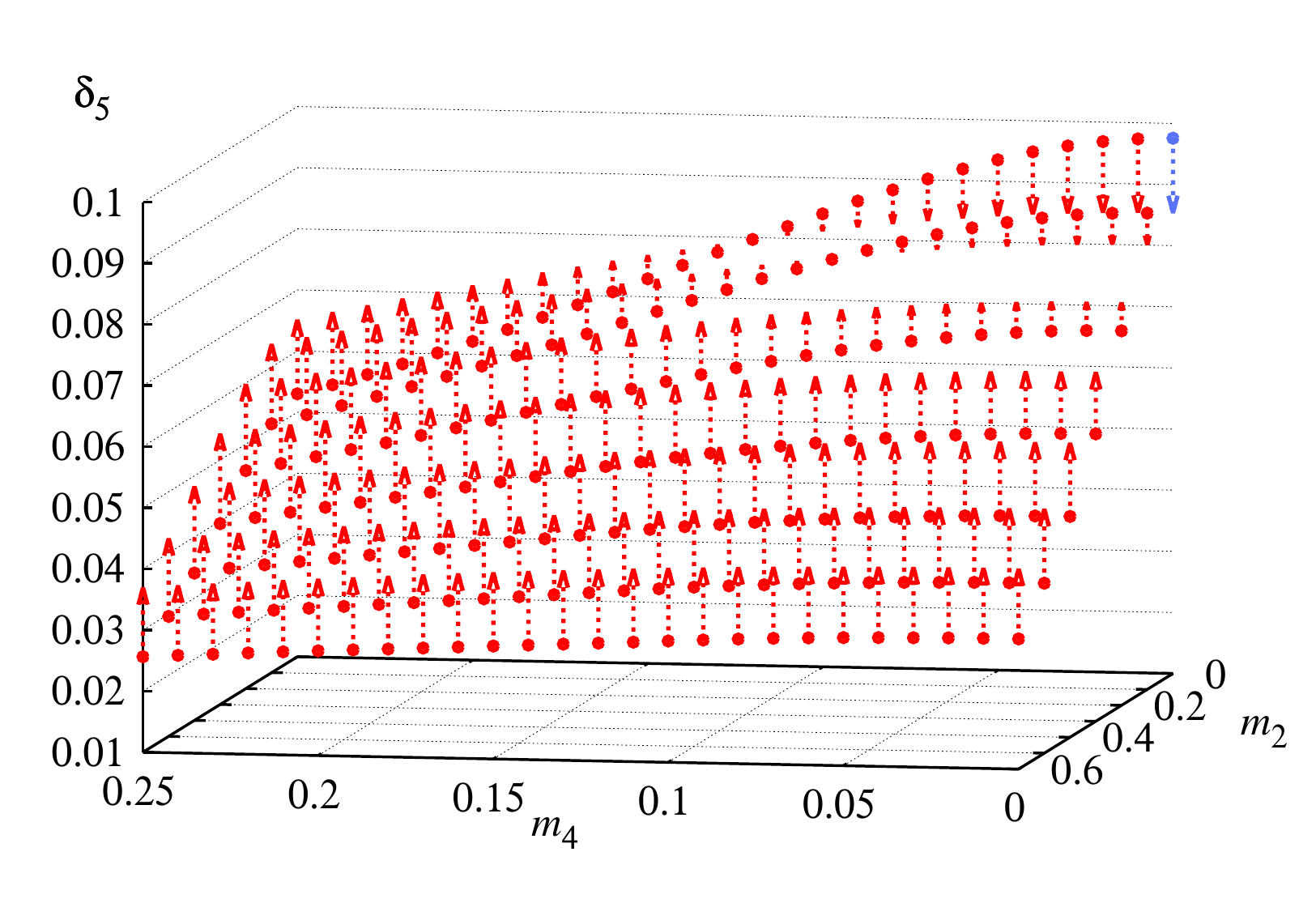}}
  \end{center}
  \caption{Relative classical binding energies $\delta_B$ for
    $B=2,3,4,5$ with pion masses turned on $m_1=1/4$. The circles are
    the relative classical binding energies $\delta_B$ without the
    inclusion of pion masses, i.e.~the data from Fig.~\ref{fig:rbe},
    whereas the heads of the arrows denote the new classical binding
    energies after inclusion of the pion mass. Blue dots and blue
    arrows denote connected isosurfaces at the half-maximum baryon
    charge level, while red dots and red arrows are disconnected. }
  \label{fig:pmrbe}
\end{figure}

We will start by computing the relative classical binding energies on
the same parameter space as used in Fig.~\ref{fig:rbe}. The color code
is used in the same way such that blue indicates a connected Skyrmion
at the level of half-maximum baryon charge isosurfaces and red
indicates a disconnected Skyrmion.
The plots in the figure are arrows from the dots (without pion
mass) to the heads of the arrows (with pion mass).
It is interesting to note that the change due to the inclusion of the
pion mass is not monotonic over the parameter space; for small
$m_2\lesssim 0.1$-$0.2$ the binding energies decrease (more drastically
for smaller values of $m_4$ than larger values), while for
$m_2\gtrsim 0.1$-$0.2$ the binding energies increase.
The same effect occurs for $m_4\sim 0.1$ and larger. 
Another feature that we can read off the figure is that the $B=3$ and
$B=4$ Skyrmions become more persistent not to deform as function of
increasing $m_2$. 
One may naively think that it may imply that smaller binding energies
may be reached before the Skyrmions split up and change their
symmetries, but the pion mass also increases the binding energies in
that region of parameter space. Therefore there are two competing
forces at play here.

\begin{figure}[!tp]
  \begin{center}
    \mbox{
      \includegraphics[width=0.49\linewidth]{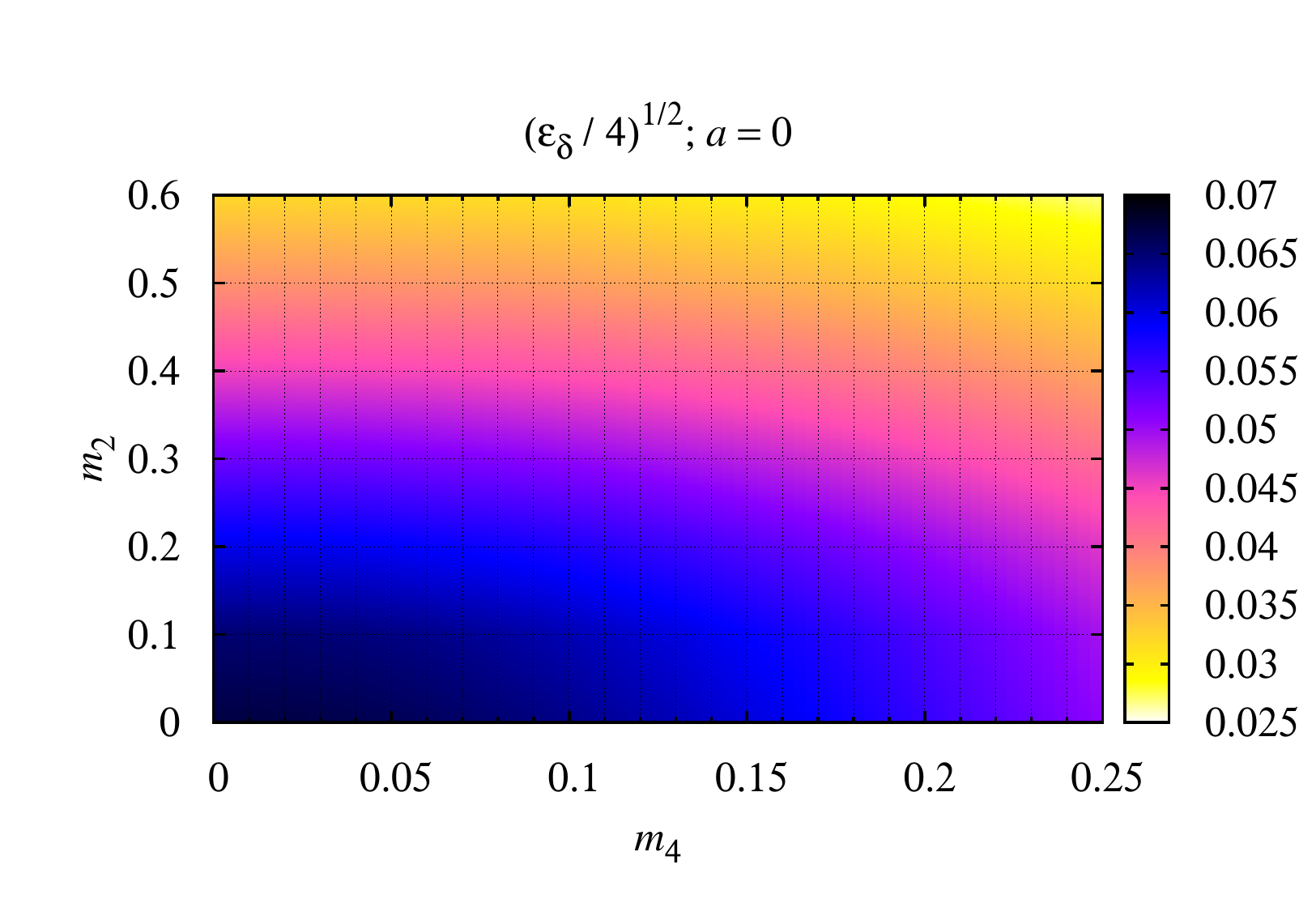}
      \includegraphics[width=0.49\linewidth]{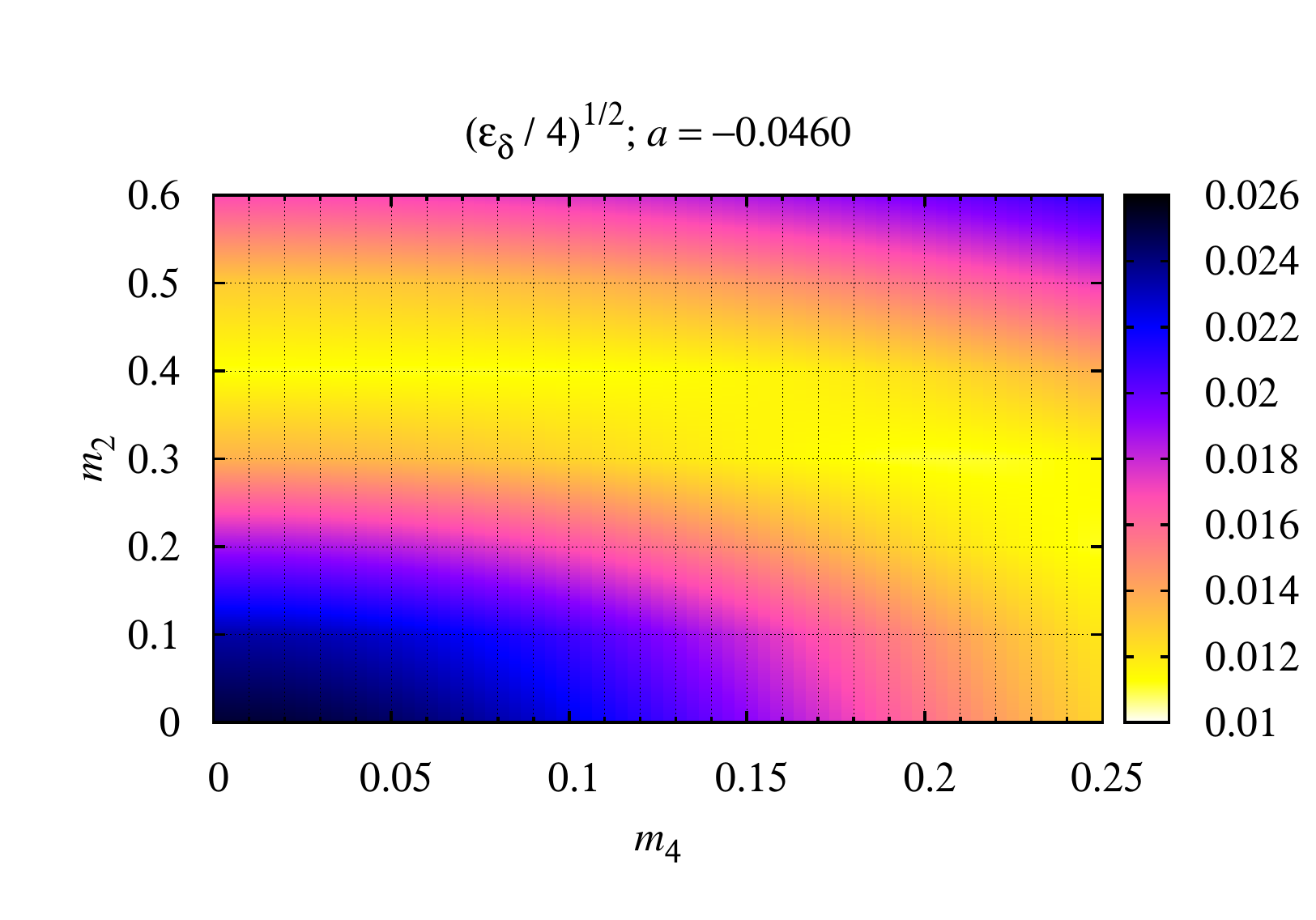}}
  \end{center}
  \caption{
    Fits of the relative classical binding energies $\delta_B$ for
    Skyrmions with pion masses $m_1=1/4$, summed up in the function
    $\varepsilon_\delta$. $\sqrt{\varepsilon_\delta/4}$ corresponds to
    the average discrepancy of binding energy, which ranges from about 7\%
    to 2.5\%. 
    The zero point, $a$ is fitted in the right panel of the figure,
    which corresponds to ignoring the $B=1$ Skyrmion's energy (shape
    fit).
  } 
  \label{fig:pmerrdelta}
\end{figure}

In Fig.~\ref{fig:pmerrdelta} we display the least-squares fit function 
$\varepsilon_\delta$ which is the average mismatch of the classical
binding energies of all the Skyrmion sectors ($B=2,3,4,5$) compared
with the experimental data.
It is seen from the figure that in this part of parameter space, the
dependence on $m_4$ is rather weak, whereas the increase of $m_2$
decreases the average classical binding energies to about 3\%.
The right-hand side panel of Fig.~\ref{fig:pmerrdelta} shows a fit to 
the shape of the binding energies ignoring the $B=1$ Skyrmion's
energy. This fit prefers points in the parameter space around
$(m_2,m_4)=(0.4,0)$ (and along a line extending in the $m_4$
direction).
This shape-fit corresponds to the situation where the $B=2,3,4,5$
Skyrmions do not receive extra contribution upon 
semi-classical quantization and the $B=1$ Skyrmion has about 4.6\%
lower energy.
Since its ground state is a spin-$\frac{1}{2}$ state, the quantum
contribution will only worsen the problem.
The quantum contribution for the spin-$\frac{1}{2}$,
isospin-$\frac{1}{2}$ state found in Ref.~\cite{Manko:2007pr} is about
2.2\% for the $B=1$ Skyrmion.

\begin{figure}[!tp]
  \begin{center}
      \includegraphics[width=0.49\linewidth]{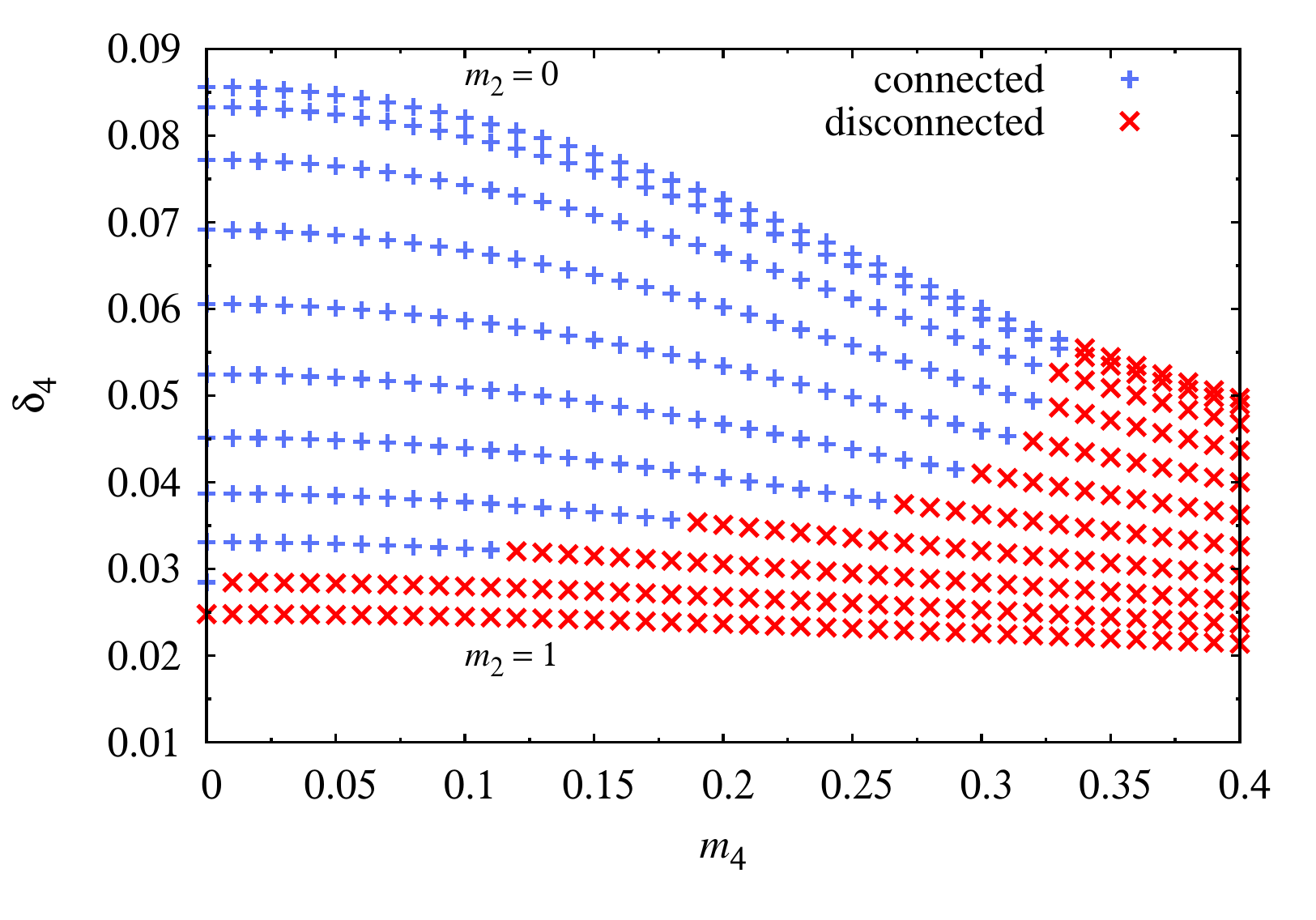}
  \end{center}
  \caption{
    Relative classical binding energies $\delta_B$ for $B=4$
    with the pion mass turned on $m_1=1/4$. The series of points is
    for $m_2=0,0.1,0.2,0.3,0.4,0.5,0.6,0.7,0.8,0.9,1$ with $m_2$
    increasing from top to bottom. The blue crosses ($+$) are 
    connected isosurfaces at half-maximum baryon charge densities
    while the red $x$s ($\times$) are disconnected. 
  }
  \label{fig:pmrbe4large}
\end{figure}

The effect of turning on the pion mass is evident in
Fig.~\ref{fig:pmrbe}, however, the parameter space is unfortunately 
too small in order to see the effect of the Skyrmions with the pion 
mass turned on, breaking up into disconnected pieces and eventually
situating themselves in an FCC lattice.
Therefore we show a larger part of the parameter space, for the $B=4$
sector in Fig.~\ref{fig:pmrbe4large}.
The situation is now quite clear.
The effect of increasing $m_4$ (from zero) is a decrease in binding
energy, but long before the binding energies of realistic nuclei are
reached, the Skyrmion breaks up into disconnected pieces and soon
prefers the FCC lattice structure.
The effect of $m_2$, on the other hand, is also a decrease in binding
energy, but much lower binding energies can be reached before the
symmetries of the Skyrmion (in the $B=4$ sector) change.
Another lesson that can be drawn from Fig.~\ref{fig:pmrbe4large} is
that once $m_2$ takes on a sizable nonzero value, then the effect of
$m_4$ is rather weak (other than breaking up the Skyrmion),
i.e.~meaning that the binding energies do not drop quickly with the
increase of $m_4$.
Due to this latter fact, we will consider only $m_4=0$ in the
remainder of the paper.

\begin{figure}[!tp]
  \begin{center}
    \mbox{
      \subfloat[]{\includegraphics[width=0.49\linewidth]{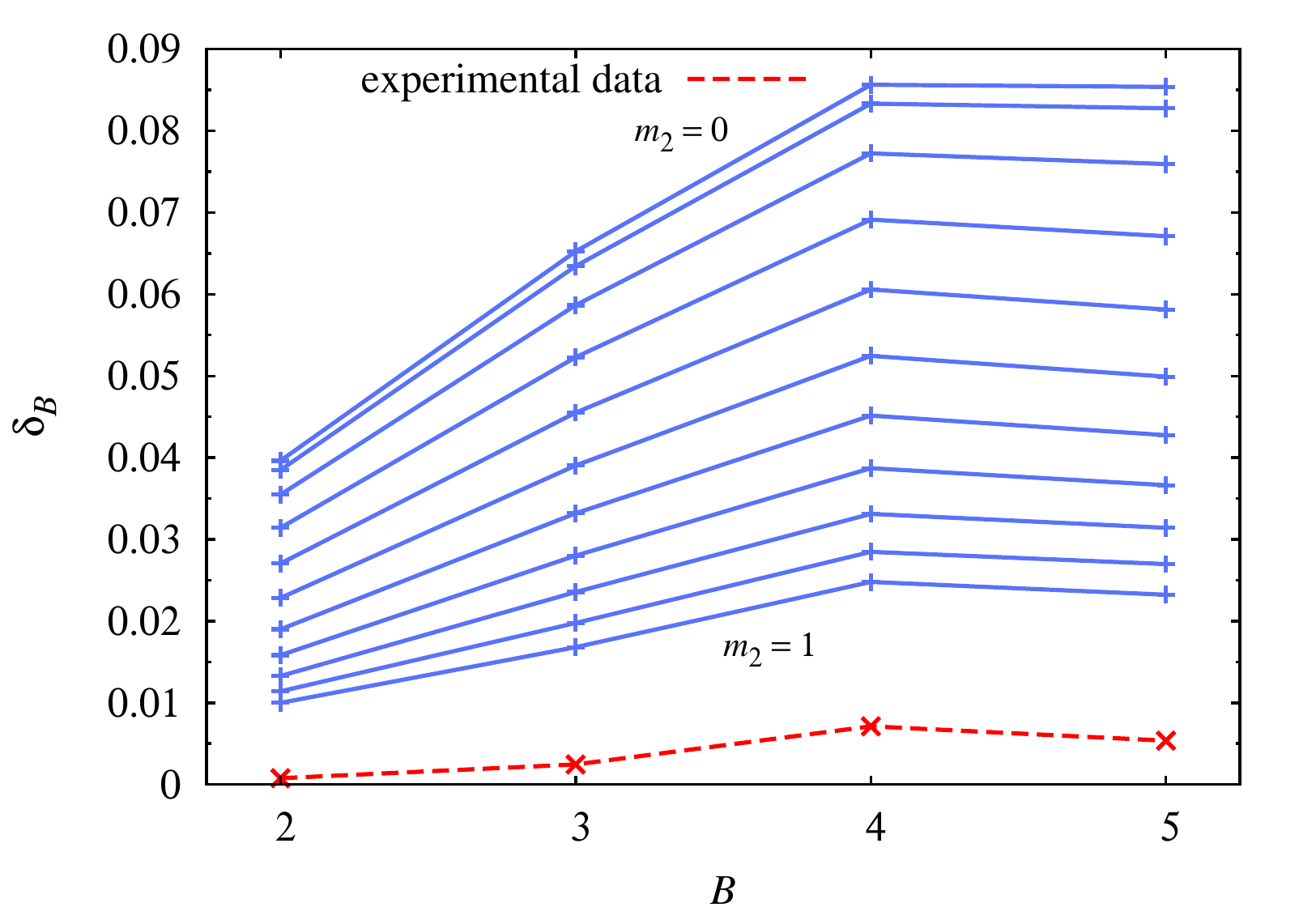}}
      \subfloat[]{\includegraphics[width=0.49\linewidth]{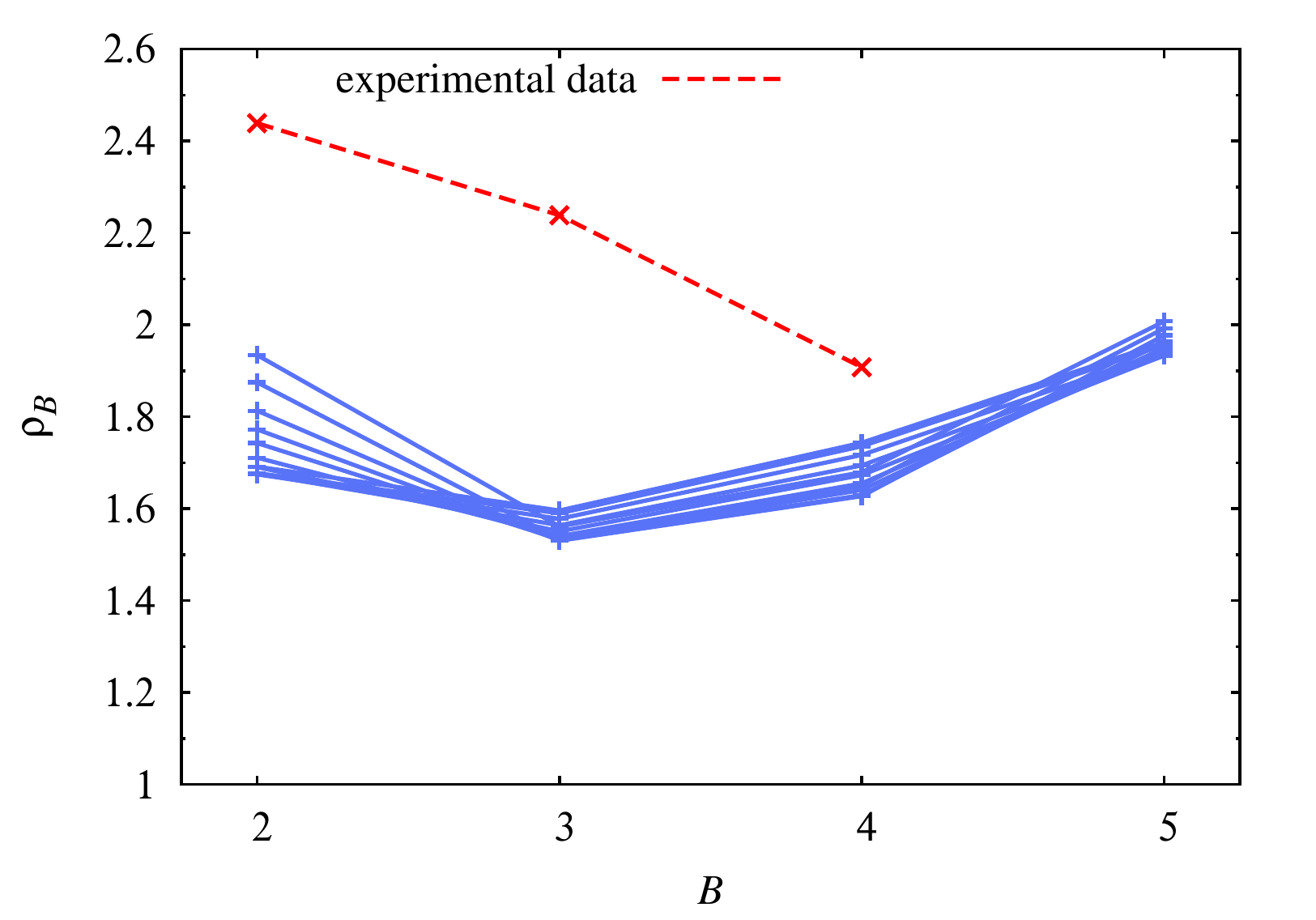}}}
  \end{center}
  \caption{
    (a) Relative classical binding energies $\delta_B$ and (b)
    relative charge radii for $B$-Skyrmions with the pion mass
    $m_1=1/4$ turned on. The series of points is for
    $m_2=0,0.1,0.2,0.3,0.4,0.5,0.6,0.7,0.8,0.9,1$ with $m_2$
    increasing from top to bottom.
    The red-dashed line is connecting the experimental data from
    (a) Tab.~\ref{tab:nuclearbinding} and (b)
    Tab.~\ref{tab:chargeradii}. 
  }
  \label{fig:pmrbeB}
\end{figure}

\begin{figure}[!tp]
  \centering
  \begin{center}
    \mbox{
      \includegraphics[width=0.2\linewidth]{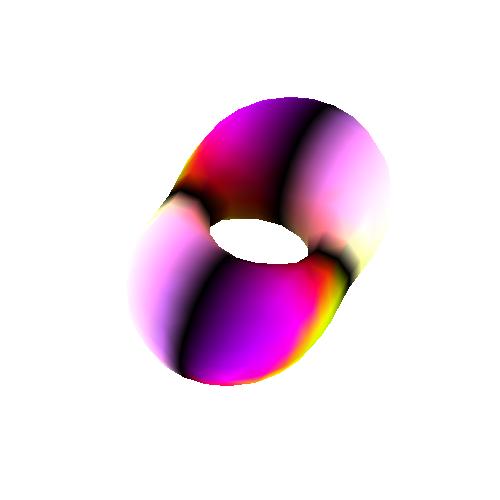}
      \includegraphics[width=0.2\linewidth]{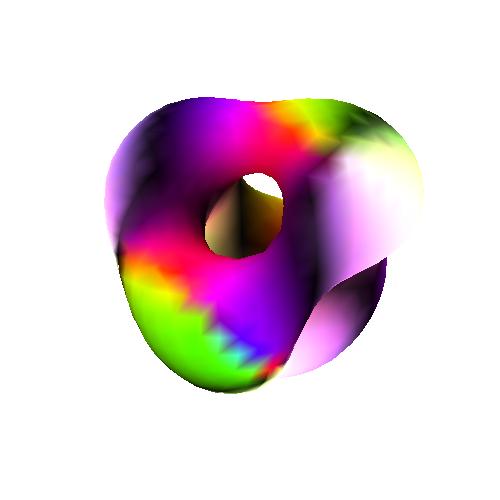}
      \includegraphics[width=0.2\linewidth]{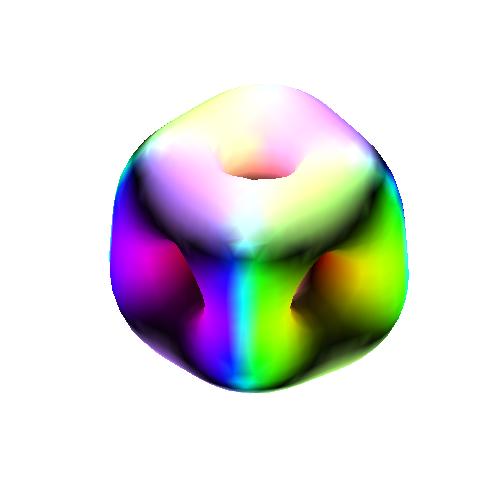}
      \includegraphics[width=0.2\linewidth]{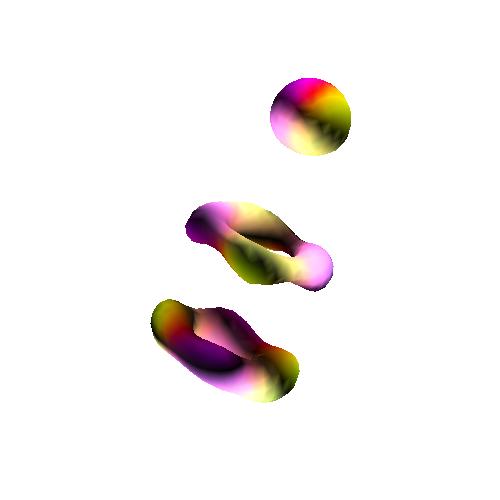}}
    \mbox{
      \includegraphics[width=0.2\linewidth]{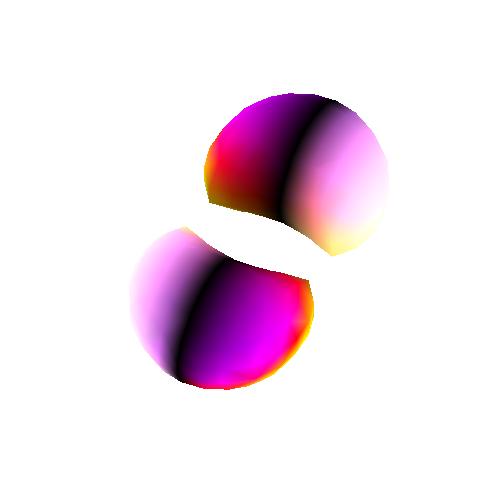}
      \includegraphics[width=0.2\linewidth]{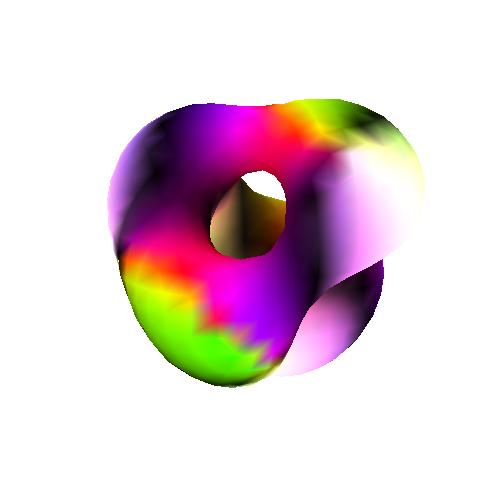}
      \includegraphics[width=0.2\linewidth]{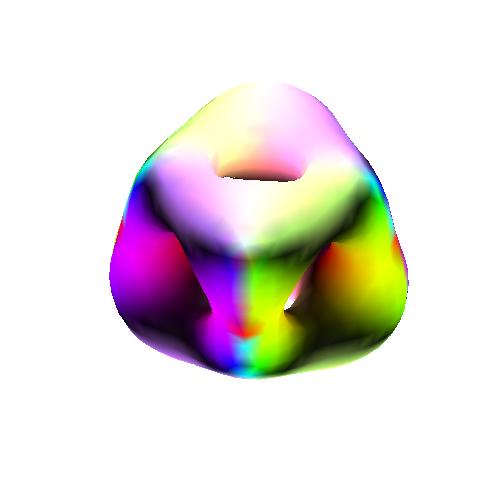}
      \includegraphics[width=0.2\linewidth]{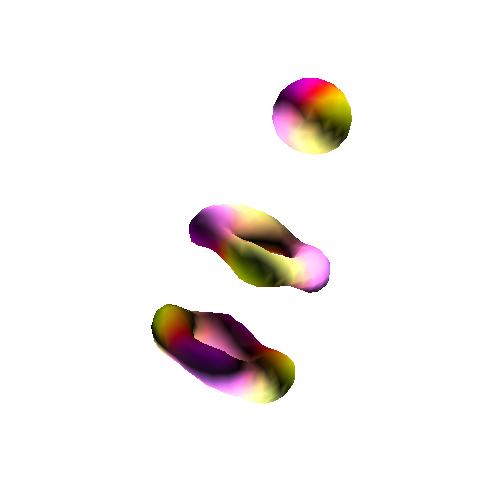}}
    \mbox{
      \includegraphics[width=0.2\linewidth]{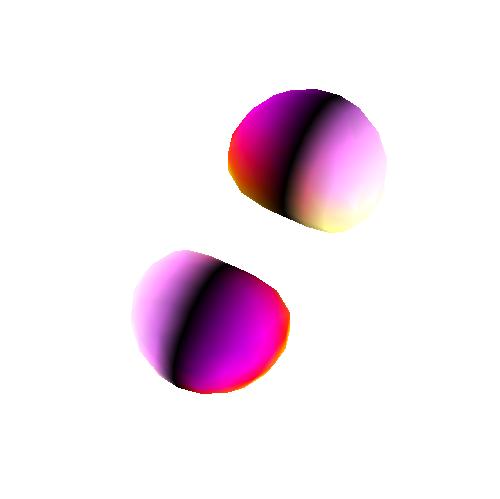}
      \includegraphics[width=0.2\linewidth]{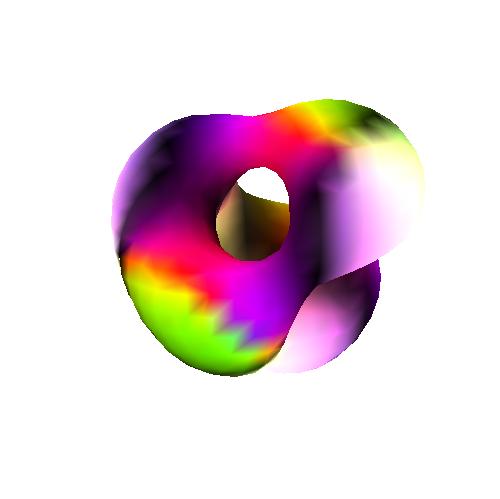}
      \includegraphics[width=0.2\linewidth]{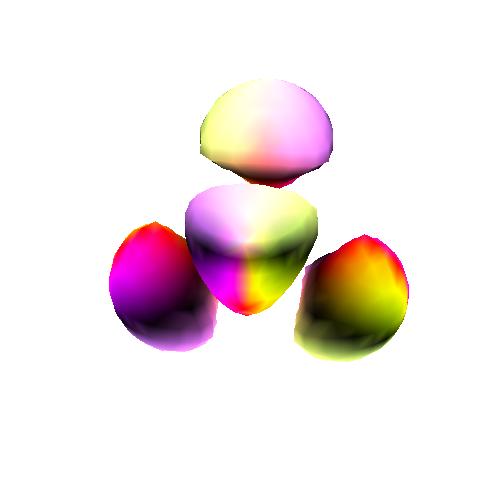}
      \includegraphics[width=0.2\linewidth]{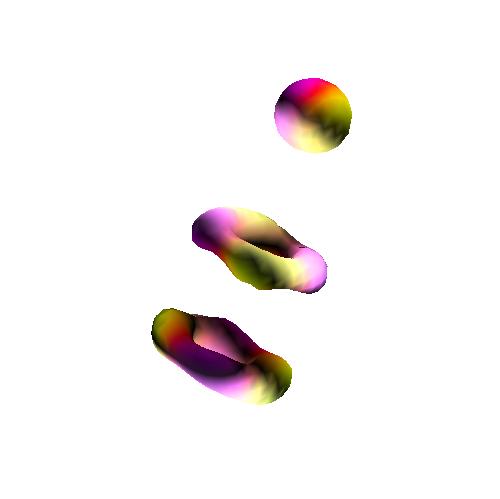}}
    \mbox{
      \includegraphics[width=0.2\linewidth]{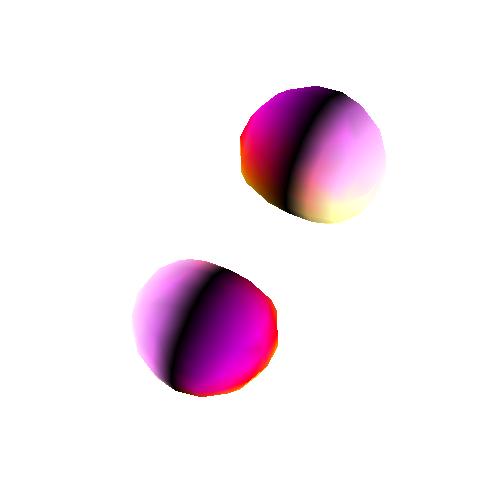}
      \includegraphics[width=0.2\linewidth]{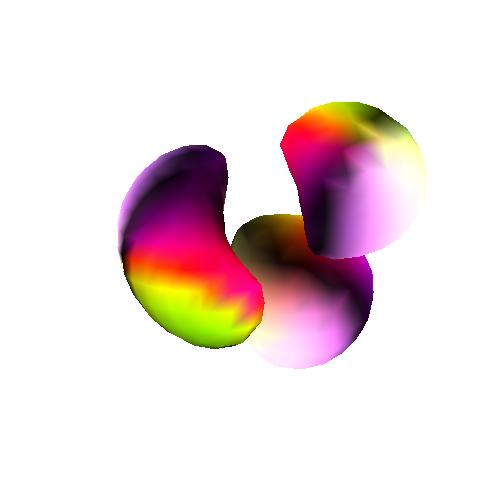}
      \includegraphics[width=0.2\linewidth]{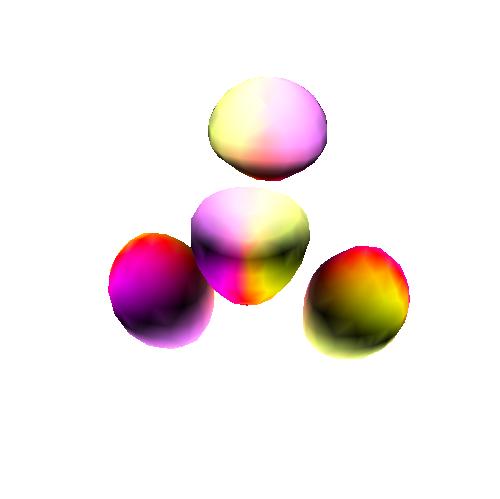}
      \includegraphics[width=0.2\linewidth]{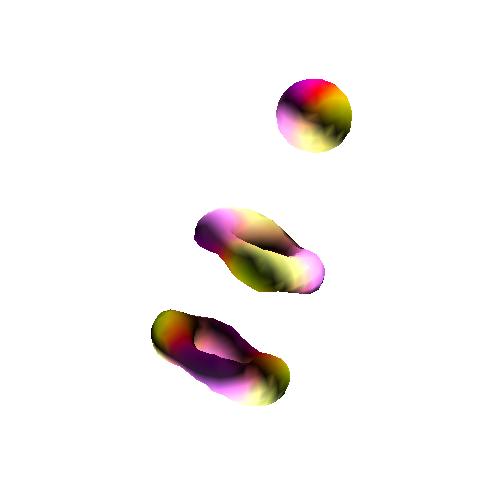}}
  \end{center}
  \caption{Isosurfaces of baryon charge density for Skyrmion solutions
  with baryon number $B=2$ through $B=5$ for $m_4=0$ as function of
  $m_2=0.7,0.8,0.9,1$ (from top to bottom).
  The coloring is described in the text. }
  \label{fig:fBpm}
\end{figure}

In Fig.~\ref{fig:pmrbeB} we consider all $B=2,3,4,5$ sectors and
display the relative classical binding energies for various values of
$m_2$ ranging from zero to one in steps of $0.1$.
The isosurfaces of their baryon charge densities at half-maximum
values are displayed in Fig.~\ref{fig:fBpm}.
It is seen from Fig.~\ref{fig:pmrbeB} that the larger the values of
$m_2$ are, the closer the classical binding energies come to those
experimentally observed. 
However, for $m_2\sim 0.7$-$0.9$ the Skyrmions start to split up into 
disconnected pieces and soon begin the transformation from platonic
symmetries to FCC lattice symmetries. 
Note that since these binding energies are purely classical binding
energies, we are not seeking an exact match between the lines of the
model calculation and the experimental data. We are merely seeking
the right ballpark value and acceptable shapes of the curves.
The experimental data for the nuclear binding energies should instead
be compared to those of the semi-classically quantized Skyrmions.
We will consider this in the next subsection.

\begin{figure}[!tp]
  \begin{center}
    \mbox{
      \includegraphics[width=0.49\linewidth]{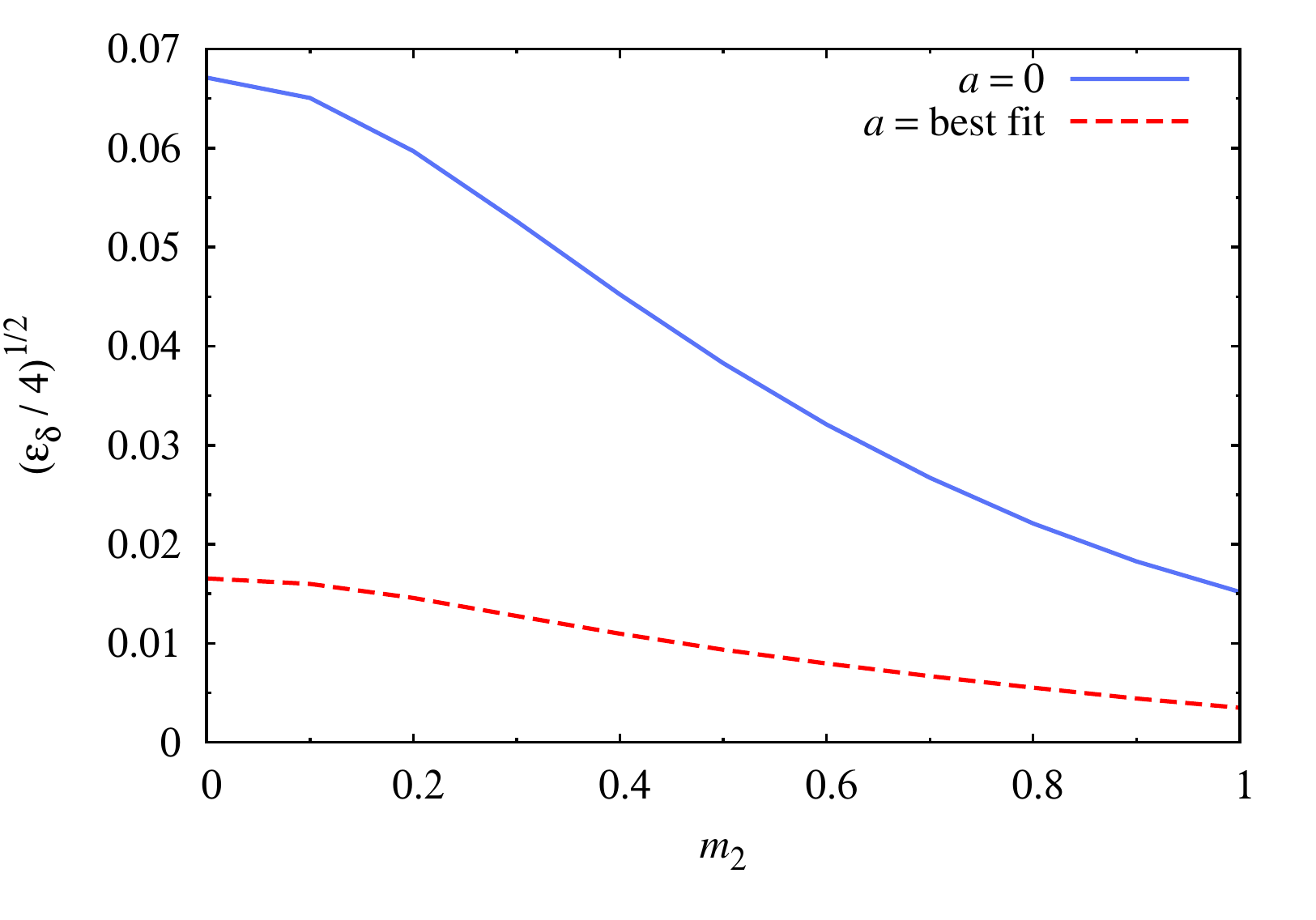}
      \includegraphics[width=0.49\linewidth]{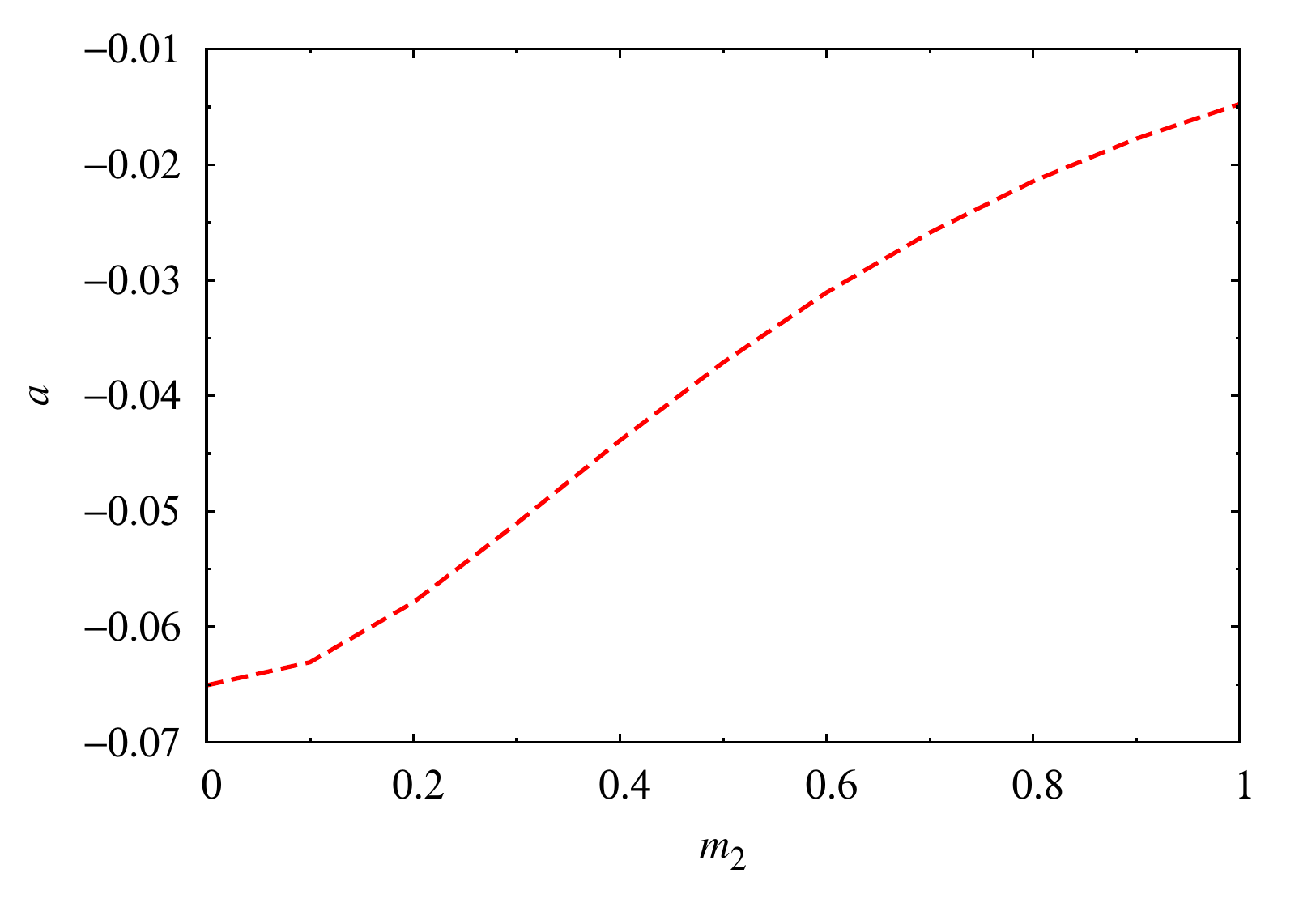}}
  \end{center}
  \caption{
    Least-squares fit of the relative classical binding energy to
    experimental data, $\sqrt{\varepsilon_\delta/4}$ as function of
    $m_2$ (left panel) for $B=2,3,4,5$ Skyrmions with the pion mass
    $m_1=1/4$ turned on.  
    $a$ is the offset constant corresponding to an extra contribution
    to the energy in the $B=1$ sector. $a$ is fitted for each value of
    $m_2$ and is shown in the right panel.  }
  \label{fig:pmerrdelta2B}
\end{figure}

Fig.~\ref{fig:pmerrdelta2B} shows the least-squares fit function
$\varepsilon_\delta$ as function of $m_2$ for Skyrmions with
$B=2,3,4,5$ and the pion mass $m_1=1/4$ turned on.
The value of $a$ that would make the model fit experimental data is
about 2-3\%, whereas the fit prefers negative values for $a$.
This means that even the classical value of the 1-Skyrmion energy is
too large by 0.5-6.5\%.

\subsection{Quantization}

We will now attempt to make a crude estimate of the semi-classically
quantized energy contributions to the Skyrmions for $m_2=0.7$, $m_4=0$
and the pion mass $m_1=1/4$ turned on.
In order to carry out a rigorous job, one should establish their
symmetries and probably not rely on the rigid body quantization
because we are working on the borderline where the Skyrmions are
trying to split up and change their symmetries.
Instead of the rigid body quantization, one should consider the
procedure carried out in Ref.~\cite{Battye:2014qva}, where the
isospinning of the Skyrmion is taken into account dynamically. This
may reveal the symmetry to be used for the quantization.
The first row of Fig.~\ref{fig:fBpm} shows the Skyrmions for $m_2=0.7$ 
and $m_4=0$. 
For the $B=2$ Skyrmion, there are two options; it may break up into
two localized (possibly deformed) spheres or it may restore axial
symmetry upon taking isospinning into account dynamically. 
The $B=3$ and $B=4$ Skyrmions retain their platonic symmetries,
namely tetrahedral and cubic symmetry, respectively.
The symmetry of the $B=5$ Skyrmion is somewhat harder to determine at
this stage.
Since we are only interested in a ballpark estimate of the
contribution from semi-classical quantization to their ground state
energies, we will (possibly unjustified) assume that they can be
quantized with the platonic symmetries used for the quantization in 
Ref.~\cite{Manko:2007pr}.
As we will see shortly, the mistake of this assumption (if wrong) will
be negligible. 

In order to add the classical Skyrmion mass and the semi-classically
quantized energy contribution, we can no longer ignore the calibration
of the model and have to make a choice.
Fitting the $B=4$ sector gives rise to
\begin{align}
\textrm{$m_2=0.7$:}\qquad
e = 3.45, \qquad
f_\pi = 69.80 \MeV, \qquad
\Rightarrow
m_\pi = 120.25 \MeV,\\
\textrm{$m_2=0.5$:}\qquad
e = 3.49, \qquad
f_\pi = 75.65 \MeV, \qquad
\Rightarrow
m_\pi = 132.14 \MeV,\\
\textrm{$m_2=0$:}\qquad
e = 3.62, \qquad
f_\pi = 88.00 \MeV, \qquad
\Rightarrow
m_\pi = 159.34 \MeV,
\end{align}
where we have used the nuclear mass of ${}^4$He: $3727 \MeV$ and the 
charge radius of ${}^4$He: 1.6755 fm.
As per usual in the Skyrme model, the physical values used in the
$B=0$ sector, i.e.~pion physics are not quite captured by the fits to
experimental nuclear data.

As can readily be seen from the above calibrations, the choice of
$m_1=1/4$ is not an accurate choice and in order to match the physical
pion mass, one should recalibrate the system for each $(m_2,m_4)$
point in the parameter space and adjust $m_1$ accordingly.
In this paper, we have merely chosen an average value that fits in the 
ballpark of the physical value. 

Using the results of Ref.~\cite{Manko:2007pr}, the semi-classical
quantum contributions to the ground state energies are given by
\begin{align}
E_1^{J=\frac{1}{2},I=\frac{1}{2}} &= \frac{f_\pi}{e} E_1 +
\frac{3e^3 f_\pi}{8V_{11}}, \\
E_2^{J=1,I=0} &= \frac{f_\pi}{e} E_2 +
\frac{e^3 f_\pi}{V_{11}}, \\
E_3^{J=\frac{1}{2},I=\frac{1}{2}} &= \frac{f_\pi}{e} E_3 +
\frac{3e^3 f_\pi}{8}\frac{U_{11} + V_{11} - 2W_{11}}{U_{11}V_{11} -
  W_{11}^2}, \\
E_4^{I=0,J=0} &= \frac{f_\pi}{e} E_4, \\
E_5^{J=\frac{3}{2},I=\frac{1}{2}} &= \frac{f_\pi}{e} E_5 +
\frac{e^3 f_\pi}{4}\frac{3U_{11} + V_{11}}{U_{11}V_{11} - W_{11}^2}
+\frac{e^3 f_\pi}{8}\frac{9U_{33} + V_{33} + 6W_{33}}{U_{33}V_{33} -
  W_{33}^2},
\end{align}
where we have restored the physical units and the tensors in our
notation are given by \cite{Manko:2007pr}
\begin{align}
  U_{ij} &= -\frac{1}{2}\int d^3x\; \Tr\left(
  c_2 T_i T_j + \frac{c_4}{4} [L_k,T_i][L_k,T_j]\right), \\
  V_{ij} &= -\frac{1}{2}\int d^3x\; \epsilon_{ilm}\epsilon_{jnp} x_l
  x_n \Tr\left(
  c_2 L_m L_p + \frac{c_4}{4} [L_k,L_m][L_k,L_p]\right), \\
  W_{ij} &= \frac{1}{2}\int d^3x\; \epsilon_{jlm} x_l \Tr\left(
  c_2 T_i L_m + \frac{c_4}{4} [L_k,T_i][L_k,L_m]\right),  
\end{align}
and $T_i\equiv\frac{i}{2}U^\dag [\tau_i,U]$.

\begin{figure}[!tp]
  \begin{center}
    \mbox{
      \subfloat[]{\includegraphics[width=0.49\linewidth]{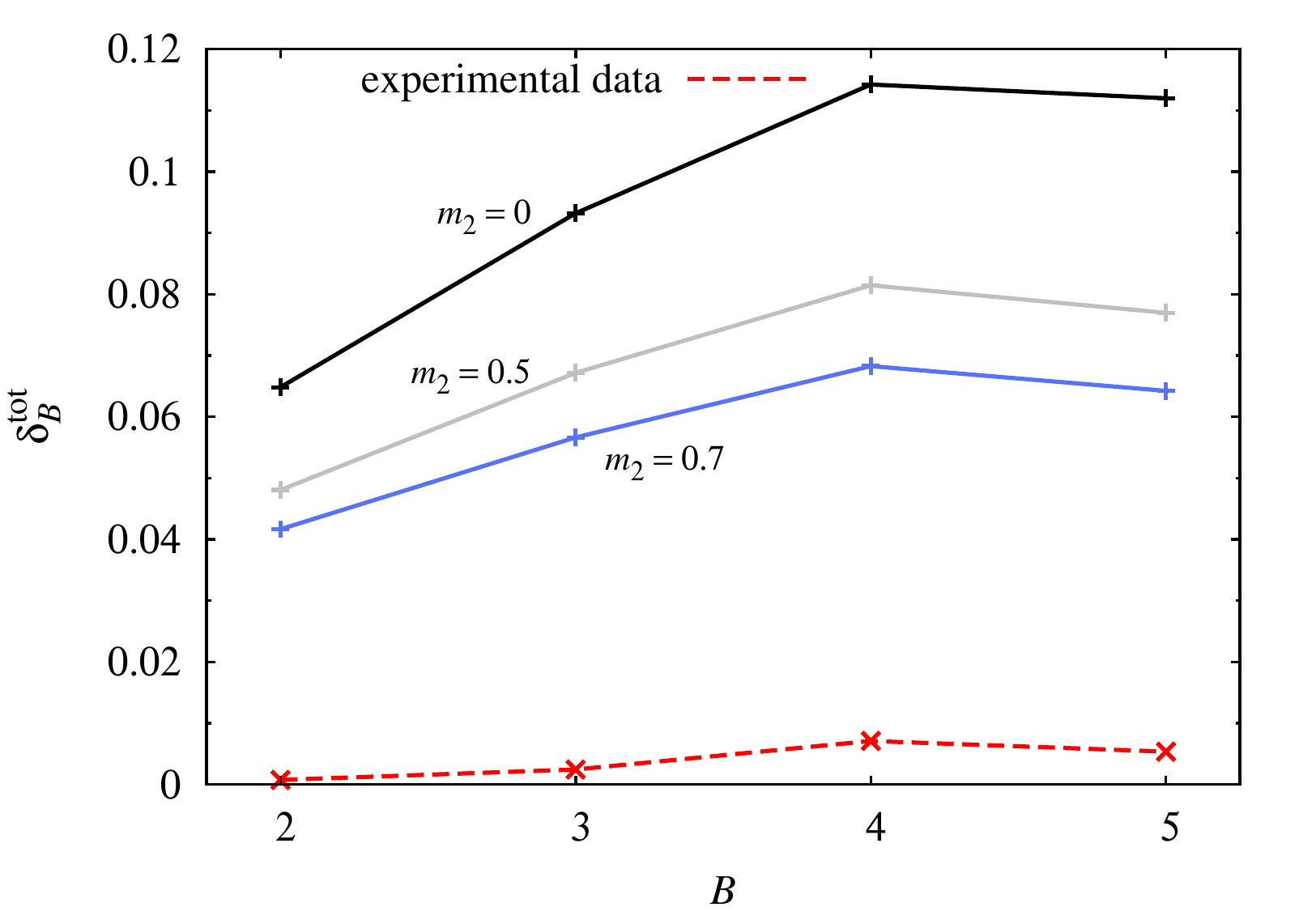}}
      \subfloat[]{\includegraphics[width=0.49\linewidth]{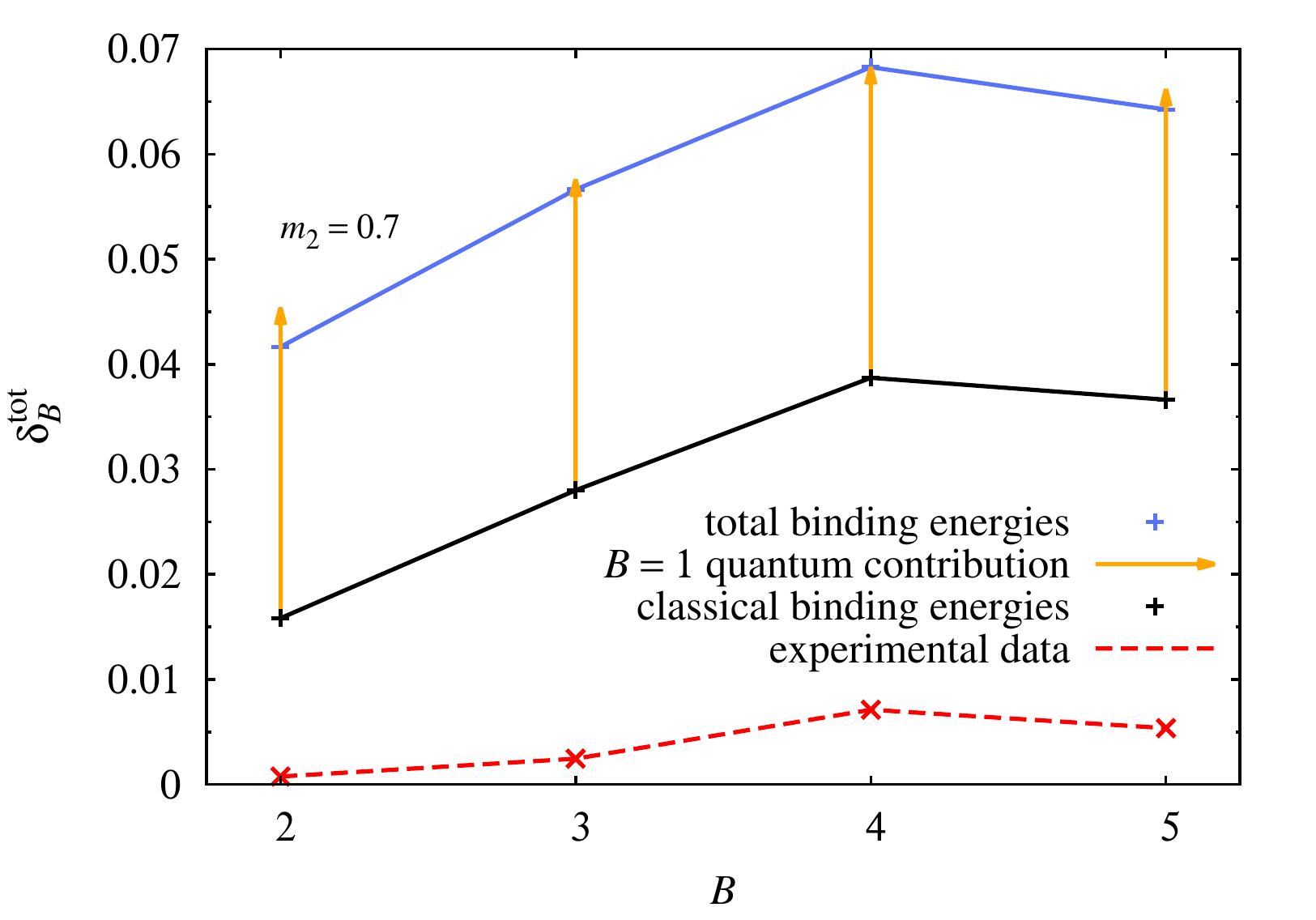}}}
  \end{center}
  \caption{(a) Relative total binding energies $\delta_B^{\rm tot}$
    with semi-classical quantum contributions from spin and isospin
    included for $B$-Skyrmions with the pion mass $m_1=1/4$ turned
    on. 
    The series of points shown is for $m_2=0,0.5,0.7$ with $m_2$
    increasing from top to bottom.
    The red-dashed line is again the experimental data from
    Tab.~\ref{tab:nuclearbinding}.
    (b) Breakdown of the semi-classical quantum contribution to the
    $m_2=0.7$ series from spin and isospin quantization. The black
    line shows the classical binding energies whereas the blue line is
    the total binding energies. The orange arrows represent the
    $B=1$ quantum contribution and the difference between the arrow
    heads and the blue line is the $B$ quantum contribution (which
    vanishes for $B=4$ as it should). }
  \label{fig:pmrbeBq}
\end{figure}

The binding energies for the quantum states -- that is, the classical
Skyrmion masses with the addition of the spin and isospin contribution
-- are shown in Fig.~\ref{fig:pmrbeBq}a for $m_2=0,0.5,0.7$, $m_4=0$
and the pion mass turned on $m_1=1/4$.
Although the $m_2=0.7$ series (roughly) retains the platonic
symmetries of the Skyrmions and lowers the binding energies till about
the 6\% level, there is still some way to go in order for the model to
reproduce the experimentally measured binding energies of nuclei.
In Fig.~\ref{fig:pmrbeBq}b is shown the breakdown of the binding
energies of the $m_2=0.7$ series.
As can be seen from the figure, now the problem of the classical
binding energies is at the same level as the quantum contributions to 
the masses.
Since the ground state of the ${}^4$He nucleus is a spin-0, isospin-0
state and the experimentally measured binding energy is
$\lesssim 1$\%, the spin contribution of almost 3\% to the $B=1$
Skyrmion energy presents an equally big problem as the classical
counterparts in the quest for low binding energies in the Skyrme
model. 
We also note that the contributions from the semi-classical
quantization to the higher-charged Skyrmions, $B=2,3,4,5$, is so low
that although they lower the binding energies, their importance is
somewhat academic at this stage.
We should remind the reader of the possibly unjustified calculation
for the spin and isospin contribution to the $B=2$ and $B=5$
Skyrmions. The proper identification of the relevant symmetries and
rigorous quantization is an interesting problem which however is
beyond the scope of this paper. 

Let us sum up what we learned so far.
The classical binding energies for the Skyrmion are generically too
large (as very well known) and the 1-Skyrmion is too small giving rise
to a large spin contribution upon semi-classical quantization (large
means 2-3\%).
The lightly bound model which uses $V_4$ as well as our new potential
$V_2$ can both decrease the classical binding energies to the level of
the experimentally observed values, but at the same time the quantum
contribution to the spin-$\frac{1}{2}$ state -- identified as the
ground state -- of the 1-Skyrmion increases and is by no means
negligible.
The lightly bound model (which uses $V_4$) cannot retain the platonic 
symmetries and lower the classical binding energies below about 5.5\%,
whereas the $V_2$ model can obtain classical binding energies below
3\%, (approximately) maintaining the platonic symmetries of the
Skyrmions.
In the $V_2$ model, the classical binding energies and the quantum
contributions are thus of the same order of magnitude, yielding
total binding energies near the 6\% level.

\section{Discussion}\label{sec:discussion}

In this paper we have studied the Skyrme model with the
addition of two scalar potentials that do not contribute to the pion
mass, but yield repulsive forces at short range -- thus reducing the
classical binding energies.
The two potentials under consideration are $V_2\propto (1-\Tr U/2)^2$
and $V_4\propto (1-\Tr U/2)^4$, where the latter was considered in
Refs.~\cite{Harland:2013rxa,Gillard:2015eia}.
Both potentials are able to lower the classical binding energies, but
$V_2$ can lower them further without breaking the platonic symmetries
well known to describe the lowest energy configurations for
$V_2=V_4=0$, i.e.~the normal Skyrme model.
Although the potential $V_2$ is able to lower the classical binding
energies to about the 3\% level, semi-classical quantization of the
1-Skyrmion -- corresponding to taking the proton or neutron spin into
account -- yields another 3\% contribution such that the total binding 
energies of the model is about 6\% -- if platonic symmetries are
wished intact.
If we give up on the platonic symmetries, both $V_2$ and $V_4$ can
lower the classical binding energies further, but since both
potentials have the effect of shrinking the 1-Skyrmion, the $V_{11}$
inertia tensor decreases, yielding an increasing spin contribution to
the energy. This thus increases the total binding energies of all the
$B$-Skyrmions (since the higher $B$ Skyrmions do not have sizable
contributions from quantization).
The question of whether the experimental values of the binding
energies can be reached with either one of the two potentials is
beyond the scope of this paper -- but an interesting future problem. 

One of the aims of this paper is to retain the platonic symmetries of
the Skyrme model, which may or may not be necessary.
The simple argument in favor of keeping the symmetries is to keep the
successes of the Skyrme model, including the description of the Hoyle
state in ${}^{12}$C \cite{Lau:2014baa}.
Further studies on this problem are however required.

It was argued in Ref.~\cite{Salmi:2014hsa} that the aloof property
that comes hand in hand with the lightly bound Skyrme model is
welcome for two reasons. The first is obviously the reduction of the
classical binding energies and the second is that the normal Skyrmions
are claimed to be too symmetric. The argument of
Ref.~\cite{Salmi:2014hsa} is based on the fact that the $B=7$ Skyrmion
fits poorly the experimentally observed data because the Skyrmion has
a very large symmetry that eliminates the states with 
spin $\tfrac{1}{2},\tfrac{3}{2}$ and $\tfrac{7}{2}$ which is in
conflict with the experimental observation that the ground state of
${}^7$Li is a spin-$\frac{3}{2}$ state.
The recent paper \cite{Halcrow:2015rvz}, however, remedies the failure
of the Skyrme model to include the spin-$\frac{3}{2}$ state by
considering quantization of the vibrational modes of the 7-Skyrmion.
The result is that a spin-$\frac{3}{2}$ state is present in the
normal Skyrme model enjoying the platonic symmetries.

Since our model does not quite achieve the requirement of very low
binding energies observed experimentally in nuclei, further
improvements are needed.
It has been observed in this paper that the pion mass term actually
increases the classical binding energies and thus exacerbates the
problem at hand.
One possibility is to switch the traditional pion mass term for
another potential also yielding the pion mass, but with different
nonlinear realization.
One candidate here is the modified pion mass term ($V_{02}$), which
was studied in
Refs.~\cite{Marleau:1990nh,Piette:1997ce,Kudryavtsev:1999zm,Kopeliovich:2005vg,Davies:2009zza,Nitta:2012wi,Gudnason:2013qba,Gudnason:2014gla,Gudnason:2014hsa}.
As discussed in Sec.~\ref{sec:model}, a large class of potentials gives
rise to the pion mass, but may have different effects on the Skyrmions
-- including their classical binding energies.

Another direction that may be considered in the search for improvement
of the model is to include the sixth-order derivative term of the BPS
Skyrme model \cite{Adam:2010fg,Adam:2010ds}.
This obviously introduces another parameter in the model, but may
yield properties that are more than welcome, for instance its near
perfect fluid properties
\cite{Adam:2014nba,Adam:2014dqa,Adam:2015lpa,Adam:2015lra}.
It has been observed in several contexts that the BPS Skyrme term
increases the size of the Skyrmion
\cite{Gudnason:2014jga,Gillard:2015eia}, which is very welcome in
light of the fact that the Skyrmions are too small and that the moment 
of inertia of the 1-Skyrmion is too small. 

One approximation that when relaxed may ameliorate the problem of the
total binding energies is the unbroken isospin symmetry. In the
setting we are working in now, the proton and the neutron are the same 
object and so the 1-Skyrmion ground state should be considered as an
average of the two. Taking the splitting in energy into account due to
the isospin breaking may improve the model. 

Finally, we have not exhausted the possibilities for potential
terms. Other powers and also non-integer powers may be considered. 
Other potentials than those we have considered may have interesting
and important effects that have not yet been explored.

The quest for finding a high-precision Skyrme-like model that can
capture important features for many (all?) nuclei is certainly
interesting and important.
We will end on this remark; whether the symmetries of the Skyrmions
should be platonic or FCC. The jury is still out.

\subsection*{Acknowledgments}

S.~B.~G.~thanks Jarah Evslin for discussions. 
S.~B.~G.~also thanks the Recruitment Program of High-end Foreign
Experts for support.

\begin{sidewaysfigure}
  \begin{center}
      \includegraphics[width=\linewidth]{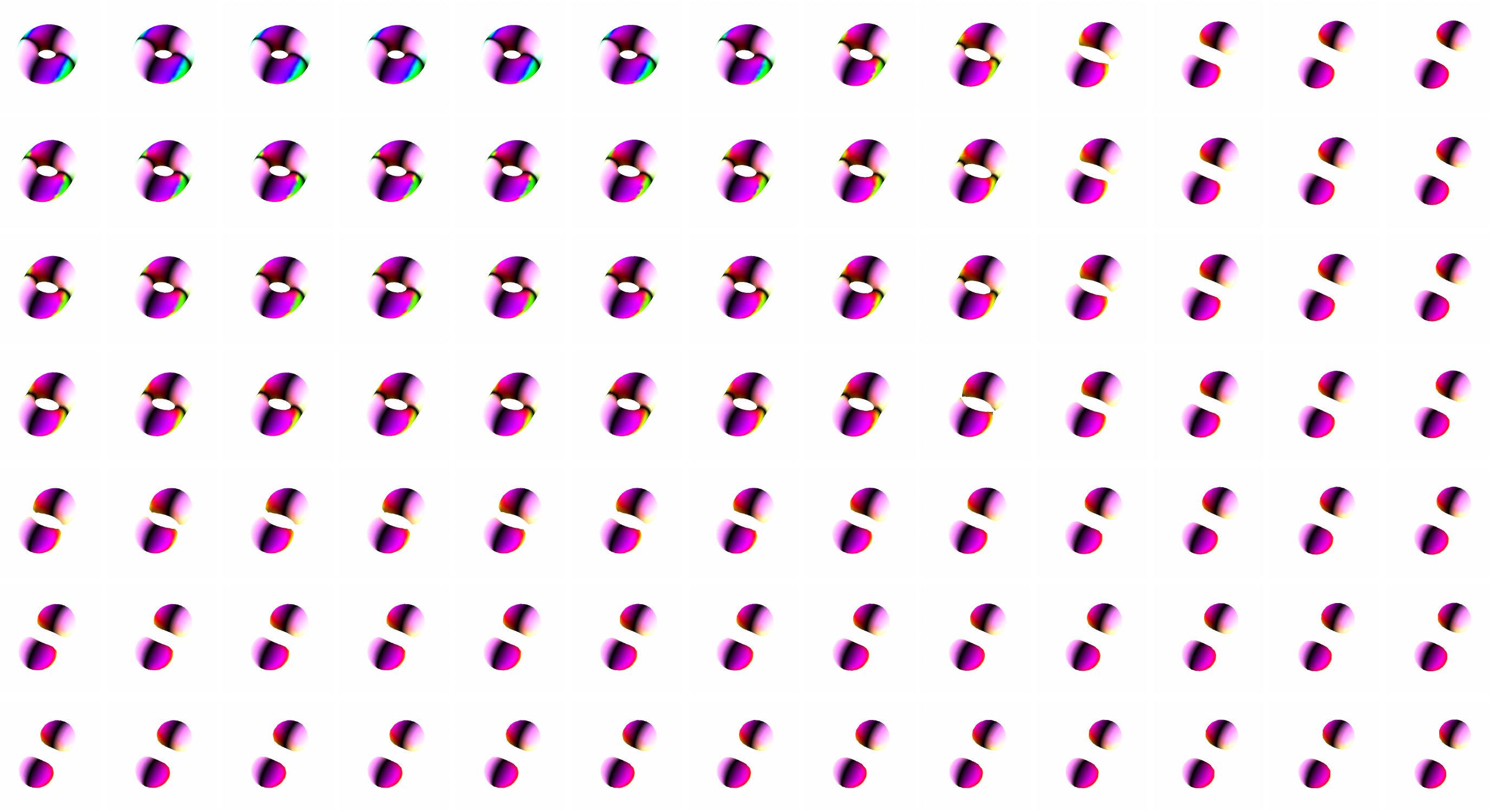}
  \end{center}
  \caption{Isosurfaces of baryon charge densities for Skyrmion solutions
  with baryon number $B=2$ in the $(m_2,m_4)$-parameter space. The
  values of
  $m_4=0,0.02,0.04,0.06,0.08,0.1,0.12,0.14,0.16,0.18,0.2,0.22,0.24$
  (increasing from left to right) while
  $m_2=0,0.1,0.2,0.3,0.4,0.5,0.6$ (increasing from top to bottom).
  The coloring is described in the text. }
  \label{fig:f2}
\end{sidewaysfigure}

\begin{sidewaysfigure}
  \begin{center}
    \includegraphics[width=\linewidth]{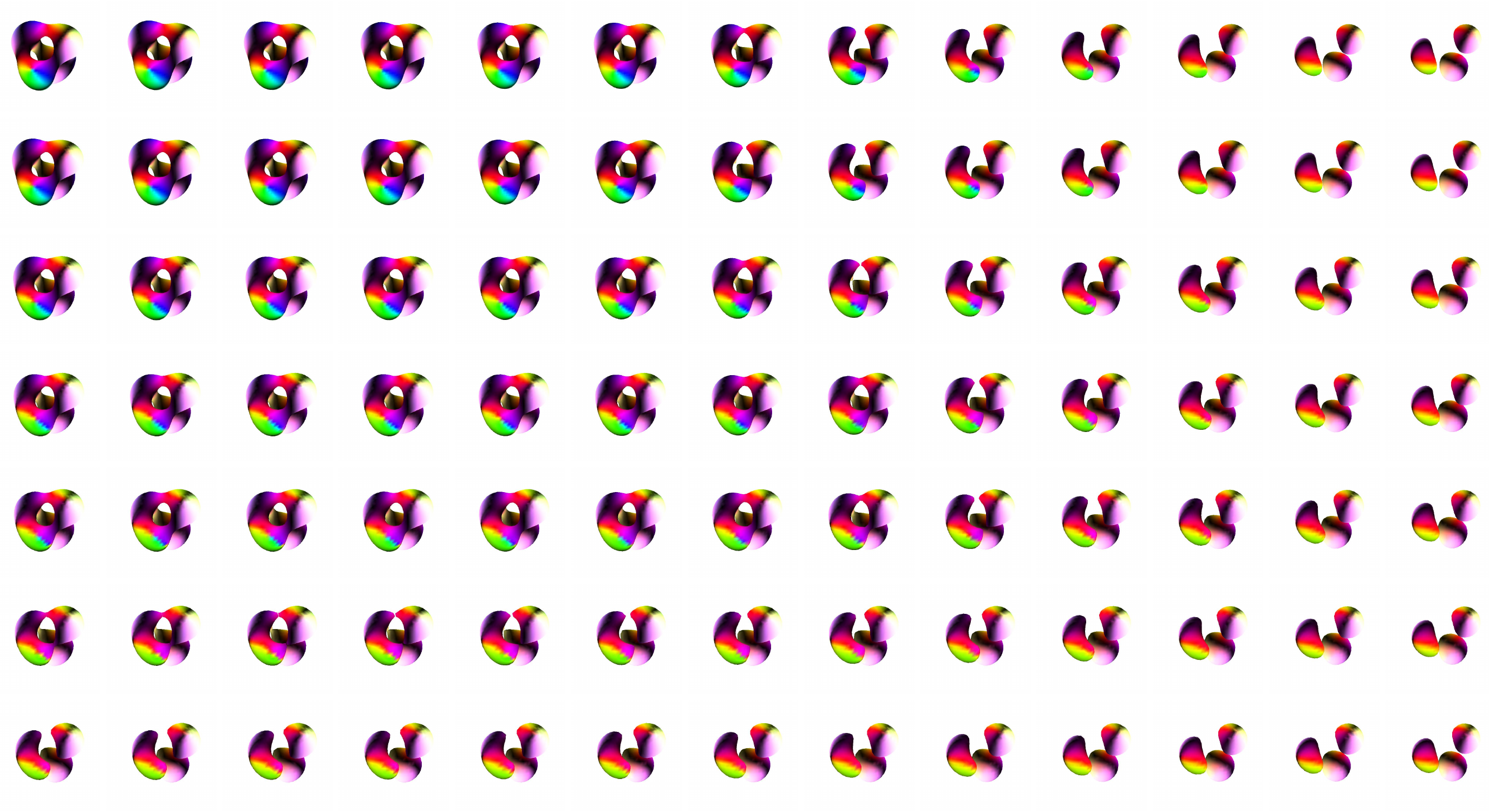}
  \end{center}
  \caption{Isosurfaces of baryon charge densities for Skyrmion solutions
  with baryon number $B=3$ in the $(m_2,m_4)$-parameter space. The
  values of
  $m_4=0,0.02,0.04,0.06,0.08,0.1,0.12,0.14,0.16,0.18,0.2,0.22,0.24$
  (increasing from left to right) while
  $m_2=0,0.1,0.2,0.3,0.4,0.5,0.6$ (increasing from top to bottom).
  The coloring is described in the text. }
  \label{fig:f3}
\end{sidewaysfigure}

\begin{sidewaysfigure}
  \begin{center}
    \includegraphics[width=\linewidth]{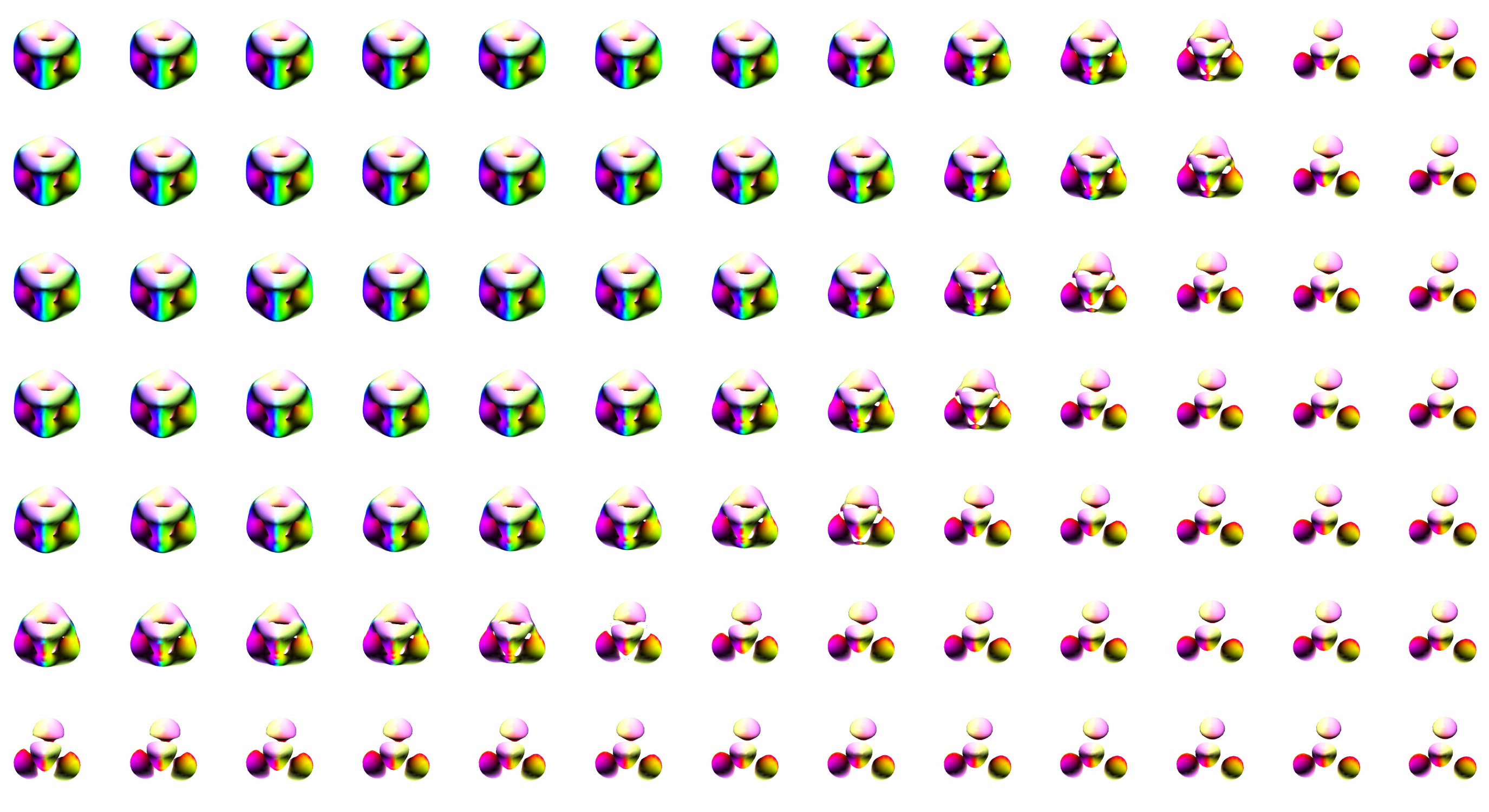}
  \end{center}
  \caption{Isosurfaces of baryon charge densities for Skyrmion solutions
  with baryon number $B=4$ in the $(m_2,m_4)$-parameter space. The
  values of
  $m_4=0,0.02,0.04,0.06,0.08,0.1,0.12,0.14,0.16,0.18,0.2,0.22,0.24$
  (increasing from left to right) while
  $m_2=0,0.1,0.2,0.3,0.4,0.5,0.6$ (increasing from top to bottom).
  The coloring is described in the text. }
  \label{fig:f4}
\end{sidewaysfigure}

\begin{sidewaysfigure}
  \begin{center}
    \includegraphics[width=\linewidth]{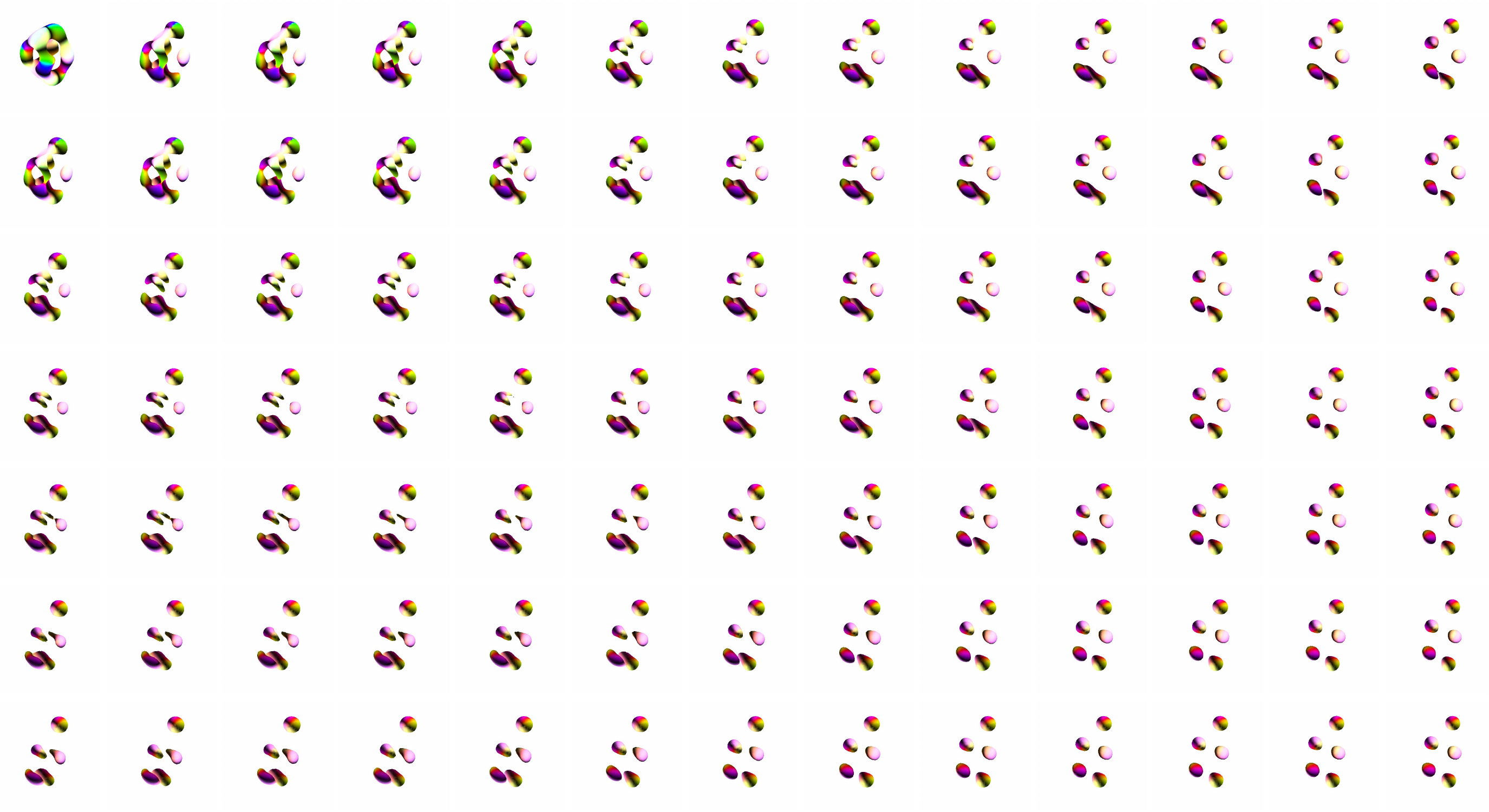}
  \end{center}
  \caption{Isosurfaces of baryon charge densities for Skyrmion solutions
  with baryon number $B=5$ in the $(m_2,m_4)$-parameter space. The
  values of
  $m_4=0,0.02,0.04,0.06,0.08,0.1,0.12,0.14,0.16,0.18,0.2,0.22,0.24$
  (increasing from left to right) while
  $m_2=0,0.1,0.2,0.3,0.4,0.5,0.6$ (increasing from top to bottom).
  The coloring is described in the text. }
  \label{fig:f5}
\end{sidewaysfigure}

\end{document}